\documentclass[10pt]{article}
\usepackage{euscript,amssymb,amsmath}
\usepackage[dvips]{graphicx}
\catcode`\@=11

\newcommand{\be}[1]{\begin{equation}\label{#1}}
\newcommand{\ee}{\end{equation}}
\newcommand{\ba}[1]{\begin{eqnarray}\label{#1}}
\newcommand{\ea}{\end{eqnarray}}
\newcommand{\rf}[1]{(\ref{#1})}
\newcommand{\nn}{\nonumber}

\begin{document}

\title{Paradoxes of dissipation-induced destabilization\\or\\
who opened Whitney's umbrella?}
\maketitle
\begin{center}
{\large Oleg N. Kirillov } \\ Dynamics and Vibrations Group, Department of Mechanical Engineering,\\
Technical University of Darmstadt,
Hochschulstr. 1, 64289 Darmstadt, Germany\\
 {\large Ferdinand Verhulst}\\ Mathematisch Instituut, University of Utrecht,\\
PO Box 80.010, 3508 TA Utrecht, the Netherlands
\end{center}
\date{}

\begin{abstract}
The paradox of destabilization of a conservative or non-conservative system by small dissipation, or Ziegler's paradox
(1952), has stimulated an ever growing interest in the sensitivity of reversible and Hamiltonian systems with
respect to dissipative perturbations. Since the last decade it has been widely accepted that dissipation-induced
instabilities are closely related to singularities arising on the stability boundary. What is less known
is that the first complete explanation of Ziegler's paradox by means of the Whitney umbrella singularity dates
back to 1956. We revisit this undeservedly forgotten pioneering result by Oene Bottema that outstripped
later findings for about half a century. We discuss subsequent developments of the perturbation analysis of dissipation-induced instabilities and applications over this period, involving structural stability of
matrices, Krein collision, Hamilton-Hopf bifurcation and related bifurcations.
\end{abstract}

\begin{flushleft}
{MSC:\\
Keywords: {\it Dissipation-induced instabilities, destabilization paradox, Ziegler's pendulum, Whitney's umbrella}}
\end{flushleft}

\section{Introduction}

`Il n'y a de nouveau que ce qui est $\rm oubli\acute{e}$'---this paraphrase of the Ecclesiastes 1:10, attributed to Marie-Antoinette, perfectly summarizes the story of the mathematical description of the destabilizing effect of vanishing dissipation in non-conservative systems.


There is a fascinating category of mechanical and physical systems which exhibit the
following paradoxical behavior: when modeled as systems without damping
they possess stable equilibria or stable steady motions, but when small
damping is introduced, some of these equilibria or steady motions become unstable.

The paradoxical effect of damping on dynamic
instability was noticed first for rotor systems which have stable steady
motions for a certain range of speed, but which become unstable when the
speed is changed to a value outside the range.

In 1879 Thomson and Tait \cite{TT79} showed that a statically unstable conservative system which has been stabilized by
gyroscopic forces could be destabilized again by the introduction of small damping
forces. More generally, they consider conservative and nonconservative linear two degrees of freedom
systems in remarkable detail. The destabilization by damping, using Routh's theorems, is implicit
in their calculations, it is not formulated as a paradox.

In 1924, to explain the destabilization of a flexible rotor in stable rotation  at a speed above the critical speed for resonance, Kimball \cite{K24} introduced a damping of the rotation, which has lead to non-conservative positional (circulatory) forces in the equations of motion of a gyroscopic system.  In 1933 Smith \cite{S33}  found that this non-conservative rotor system loses stability when the speed of rotation $\Omega>\omega\left(1+\frac{\delta}{\nu}\right)$, where $\omega$ is the undamped natural whirling frequency (the critical speed for
resonance) and $\delta$ and $\nu$ are the viscous damping constants for the stationary and
rotating damping mechanisms. In Smith's model, the destabilizing effect of the damping
of rotation $(\nu)$, observed also by Kapitsa \cite{K39}, was compensated by the stationary damping $(\delta)$. This was a first demonstration of a strong influence of
the spatial distribution of damping (or equivalently the modal distribution) on the borderlines between stability and instability domains in multi-modal non-conservative systems \cite{C95,Zhu92}.

In the 1950s and 1960s the publications of Ziegler \cite{Zi52,Zi53}, Bolotin \cite{Bo63,BZ69}, and Herrmann \cite{HJ65,H67}, motivated by aerodynamics applications, initiated a considerable activity in the
investigation of dynamic instability of equilibrium configurations of structures under non-conservative loads.
The canonical problem was the flutter of a vertical flexible cantilever column under a compressive
non-conservative or follower load which remains tangent to the bending column.
In the flutter mode the tip of the
column is preponderantly slanted towards the left during the half-cycle in
which the tip is moving towards the right and vice versa in the following
half-cycle. This snake-like oscillation permits the follower force to do
positive work on each cycle \cite{C95}.

The strong influence of the
spatial or modal distribution of damping within the structure on its stability under  non-conservative loading,
observed in these publications, should not have been surprising in the light of earlier findings of rotor dynamists.
However, they revealed explicitly the most dramatic and
paradoxical aspect of the sensitivity of the stability of the nonconservative structures to small damping forces.
It turned out that the critical load for
a structure with small damping may be considerably smaller than that for
the same structure without damping. In other words, there is a wide range
of loads for which the undamped structure is stable, but which produce
instability as soon as a tiny bit of damping is added to the structure.

These phenomena were actively studied in the 1960's to provide more basic
understanding and they have continued to be studied with more sophisticated tools, including early attempts
to employ singularity theory \cite{TZ81},
 until in the mid 1990s it was understood \cite{HR95,S96} that
the destabilization paradox is related to the Whitney umbrella singularity of the stability boundary \cite{W43,W44}.
After describing in the first sections Whitney's umbrella and Ziegler's paradox,
we make in section \ref{Bot} a sharp turn to the 1950s to revisit an article of Oene Bottema \cite{B56},
 who in 1956 first made this discovery and clarified the paradox.
Surprisingly, this paper surpassed the attention of most scientists during five decades.

In section \ref{Wh}  we will relate these results to singularity theory,
in sections \ref{HmH} and \ref{app} we show in various ways their extension to  finite- and
 infinite-dimensional systems  using
perturbation theory of multiple eigenvalues, in section \ref{Par} we focus on periodic systems, and
in the remainder we discuss applications in physics and engineering.

\section{Whitney's umbrella}
\label{Wh}

\begin{figure}[h]
\includegraphics[width=0.8\textwidth]{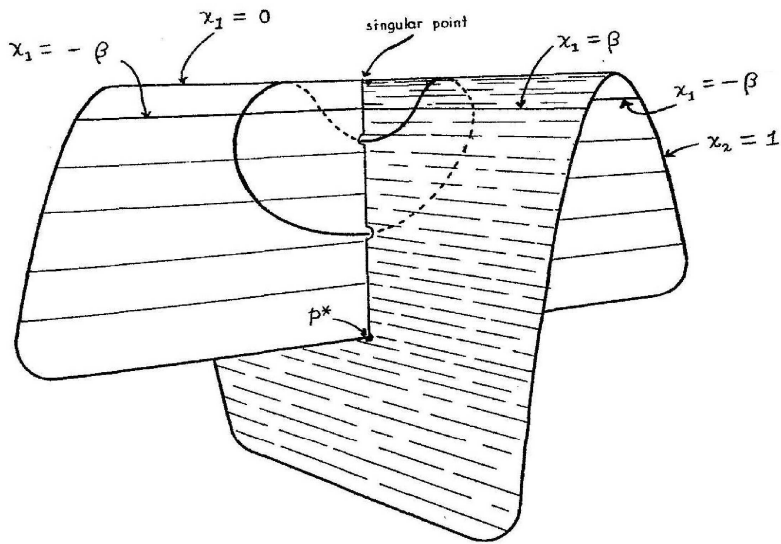}
\caption{\label{fig0} Whitney's original sketch of the umbrella \cite{W44}.}
\end{figure}

In a remarkable paper of 1943 \cite{W43}, Hassler Whitney described singularities of maps of a differential
$n$-manifold into $E^m$ with $m= 2n-1$. It turns out that in this case a special kind of
singularity plays a prominent role. Later, the local geometric structure of the manifold
near the singularity has been aptly called `Whitney's umbrella'. In Fig.~\ref{fig0} we reproduce the original sketch of the singular surface
from the companion article \cite{W44}.

The paper contains two main theorems. Consider the $C^k$ map $f: E^n \mapsto E^m$ with
$m= 2n-1$.
\begin{enumerate}
\item The map $f$ can be altered slightly, forming $f^*$, for which the singular points are
isolated. For each such an isolated singular point $p$, a technical regularity condition $C$ is valid
which relates to the map $f^*$ of the independent vectors near $p$ and of the differentials, the vectors
in tangent space.
\item Consider the map $f^*$ which satisfies condition $C$. Then we can choose coordinates
$x= (x_1, x_2, \cdots, x_n)$ in a neighborhood of $p$ and coordinates $y= (y_1, y_2, \cdots, y_m) $
(with $m= 2n-1$) in a neighborhood of $y=f(p)$ such that in a neighborhood of $f^*(p)$ we have
exactly
\begin{eqnarray*}
y_1 &=& x_1^2,\\
y_i &=& x_i,\,\,\,i=2, \cdots, n,\\
y_{n+i-1} &=& x_1x_i,\,\,\,i=2, \cdots, n.
\end{eqnarray*}
\end{enumerate}

If for instance $n=2,\, m=3$, the simplest interesting case, we have near the origin
\begin{equation}
y_1 = x_1^2,\,\,y_2 = x_2,\,\,y_3 = x_1x_2,
\end{equation}
so that $y_1 \geq 0$ and on eliminating $x_1$ and $x_2$:
\be{w} y_1y_2^2 - y_3^2 =0. \ee
Starting on the $y_2$-axis for $y_1=y_3=0$, the surface widens up for increasing values of $y_1$.
For each $y_2$, the cross-section is a parabola;
as $y_2$ passes through $0$, the parabola degenerates to a half-ray, and opens
out again (with sense reversed); see Fig.~\ref{fig0}.

Note that because of the $C^k$ assumption for the differentiable map $f$, the analysis is delicate.
There is a considerable simplification of the treatment if the map is analytical.

The analysis of singularities of functions and maps is a fundamental ingredient for bifurcation studies
of differential equations. After the pioneering work of Hassler Whitney and Marston Morse, it has become
a huge research field, both in theoretical investigations and in applications. We can not even present
a summary of this field here, so we restrict ourselves to citing a number of survey texts and
discussing a few key concepts and examples. In particular we mention \cite{Ar71}, \cite{GS85}, \cite{GS88}, \cite{Ar83},
\cite{AA88} and \cite{AA93}. A monograph relating bifurcation theory with normal forms and
numerics is \cite{Kuz04}.

The relation between singularities of functions and critical points or equilibria of differential equations
becomes relatively simple when considering Hamiltonian and gradient systems. Consider for instance the time-independent
Hamiltonian function $H({\bf p}, {\bf q})$ with ${\bf p}, {\bf q} \in {\mathbb{R}}^n$. Singularities of the function $H$ are found
in the set $ {\mathbb{R}}^{2n}$ where
\[ \frac{\partial H}{\partial {\bf p}} = \frac{\partial H}{\partial {\bf q}} = 0. \]
These points correspond with the critical points (equilibria) of the Hamiltonian equations of motion
\[  \dot{\bf p} = \frac{\partial H}{\partial {\bf q}},\,\,\dot{\bf q} = - \frac{\partial H}{\partial {\bf p}}. \]
More in general, consider the dynamical system described by the autonomous ODE
\[ \dot{\bf x} = {\bf f}({\bf x}),\,\,{\bf x} \in {\mathbb{R}}^n,\,\,{\bf f}:  {\mathbb{R}}^n \mapsto {\mathbb{R}}^n. \]
An equilibrium ${\bf x}_0$ of the system arises if ${\bf f}({\bf x}_0)=0$. With a little smoothness of the
map ${\bf f}$ we can linearize near ${\bf x}_0$ so that we can write
\begin{equation}
\dot{\bf x} = {\bf A}({\bf x}-{\bf x}_0) + {\bf g}({\bf x})
\label{lin}
\end{equation}
with ${\bf A}$ a constant $n \times n -$ matrix, ${\bf g}({\bf x})$ contains higher-order terms only. In other words
\[ \lim_{{\bf x} \rightarrow {\bf x}_0}  \frac{\| {\bf g}({\bf x}) \|}{\| {\bf x} - {\bf x}_0 \|} = 0, \]
${\bf g}({\bf x})$ is tangent to the linear map in ${\bf x}_0$.

The properties of the matrix ${\bf A}$ determine in a large number of cases the local behavior of the dynamical system.
In a seminal paper \cite{Ar71}, Arnold considers families of matrices, smoothly depending on a number
of parameters (denoted by vector ${\bf p}$). So, for the constant $n \times n -$ matrix we write
${\bf A}_{\bf p}$. Suppose that for ${\bf p} =0$, ${\bf A}_0$ is in Jordan normal form. Choosing
${\bf p}$ in a neighborhood of ${\bf p} =0$ produces a {\em deformation} (or perturbation)
of ${\bf A}_0$, assuming that near ${\bf p} =0$ the entries of ${\bf A}_{\bf p}$ can be expanded in a
convergent power series in the parameters. A deformation is {\em versal} if all other deformations
near ${\bf p} =0$ are equivalent under smooth change of parameters.

The paper \cite{Ar71} uses normal forms to obtain suitable versal deformations. These are associated
with the bifurcations of the linearized system (\ref{lin}). Note that although a matrix induces a linear
map, the corresponding eigenvalue problem produces a nonlinear characteristic equation. In addition,
the parameters involved, make it necessary to analyze maps of $ {\mathbb{R}}^{n}$ into $ {\mathbb{R}}^{m}$.
For instance in the following sections we meet with maps from $ {\mathbb{R}}^{2}$ into  ${\mathbb{R}}^{3}$ as studied by Whitney \cite{W43}. Nevertheless, in 1943 it was hard to imagine that this study of global analysis, a pure mathematical abstraction, would find already an industrial application in the next decade.

\section{Ziegler's paradox}

In 1952 Hans Ziegler of ETH Zurich published a paper \cite{Zi52} that became classical and widely known
in the community of mechanical engineers; it also attracted the attention of mathematicians.
Studying a simplified two-dimensional model of an elastic rod, fixed at one end and compressed by a tangential end load, he
unexpectedly encountered a phenomenon with a paradoxal character: the domain of stability of the Ziegler's pendulum changes
in a discontinuous way when one passes from the case in which the damping is very small to that where it  has vanished \cite{Zi52,Zi53}.

\begin{figure}[h]
\includegraphics[width=0.8\textwidth]{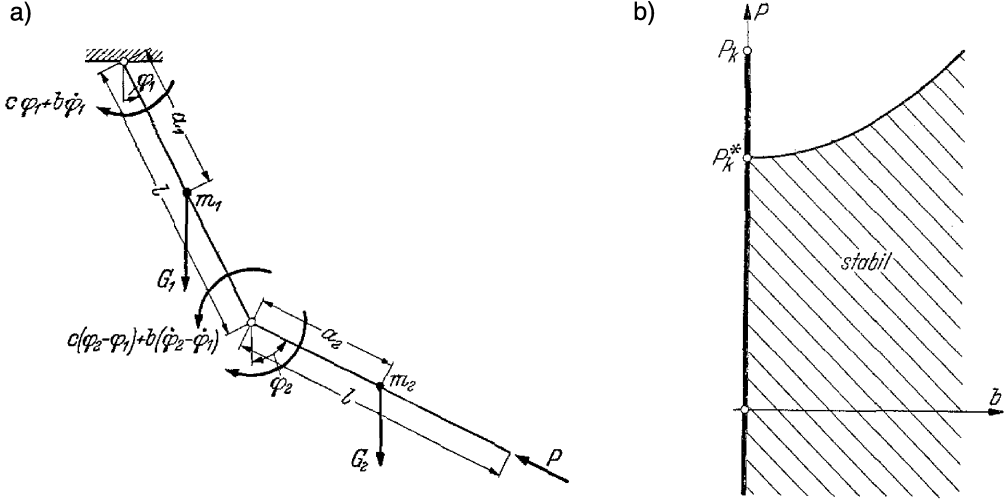}
\caption{\label{fig1} Original drawings from the Ziegler's work of 1952 \cite{Zi52}: (a) double linked pendulum under the follower load, (b) (bold line) stability interval of the undamped pendulum and (shaded area) the domain of asymptotic stability of the damped pendulum with equal coefficients of dissipation in both joints.
 If $b=0$ we have no dissipation and stability for follower force $P$ if $P < P_k$.}
\end{figure}

Ziegler's double pendulum presented in Fig.~\ref{fig1}(a) consists of two rigid rods of length $l$ each, whose inclinations with respect to the vertical are denoted as $\varphi_1$ and $\varphi_2$. Two masses $m_1$ and $m_2$ with the weights $G_1$ and $G_2$ are concentrated at the distances $a_1$ and $a_2$ from the joints. The elastic restoring torques and the damping torques at the joints are $c\varphi_1$, $c(\varphi_2-\varphi_1)$ and $b_1\dot\varphi_1$, $b_2(\dot\varphi_2-\dot\varphi_1)$, respectively.
With these assumptions the kinetic energy of the system is
\be{z1}
T=\frac{1}{2}\left[ (m_1a_1^2+m_2l^2)\dot\varphi_1^2+2m_2la_2\dot\varphi_1\dot\varphi_2+m_2a_2^2\dot\varphi_2^2\right],
\ee
while the potential energy reads
\be{z2}
V=\frac{1}{2}\left[(G_1a_1+G_2l+2c)\varphi_1^2-2c\varphi_1\varphi_2+(G_2a_2+c)\varphi_2^2 \right].
\ee
The generalized dissipative and non-conservative forces are then
\be{z3}
Q_1=Pl(\varphi_1-\varphi_2)-((b_1+b_2)\dot\varphi_1-b_2\dot\varphi_2),\quad Q_2=b_2(\dot\varphi_1-\dot\varphi_2).
\ee
Writing the Lagrange's equations of motion $\dot L_{\dot\varphi_i}-L_{\varphi_i}=Q_i$, where $L=T-V$ and
a dot denotes time differentiation, and assuming $G_1=0$ and $G_2=0$ for simplicity, we find
\be{z4}
\left(
  \begin{array}{rr}
    m_1 a_1^2+m_2l^2 & m_2 l a_2 \\
    m_2la_2 & m_2 a_2^2 \\
  \end{array}
\right)\left(
         \begin{array}{c}
           \ddot \varphi_1 \\
           \ddot \varphi_2 \\
         \end{array}
       \right)+
       \left(
  \begin{array}{rr}
    b_1+b_2 & -b_2 \\
    -b_2 & b_2 \\
  \end{array}
\right)\left(
         \begin{array}{c}
           \dot \varphi_1 \\
           \dot \varphi_2 \\
         \end{array}
       \right)+
              \left(
  \begin{array}{rr}
    -Pl+2c & Pl-c \\
    -c & c \\
  \end{array}
\right)\left(
         \begin{array}{c}
           \varphi_1 \\
           \varphi_2 \\
         \end{array}
       \right)=0.
\ee
With the substitution  $\varphi_i=A_i exp(\lambda t)$, equation \rf{z4} yields a
$4$-dimensional eigenvalue problem with respect to the spectral parameter $\lambda$.

Putting $m_1=2m$, $m_2=m$, $a_1=a_2=l$, $b_1=b_2=b$ and assuming that dissipation is absent $(b=0)$, Ziegler found from the characteristic equation that the vertical equilibrium position of the pendulum looses its stability when the magnitude of the follower force exceeds the critical value $P_k$, where
\be{z5}
P_k=\left(\frac{7}{2}-\sqrt{2} \right)\frac{c}{l}\simeq2.086\frac{c}{l}.
\ee

In the presence of damping $(b > 0)$ the Routh-Hurwitz condition yields the new critical follower load
that depends on the square of the damping coefficient $b$
\be{z6}
P_k(b)=\frac{41}{28}\frac{c}{l}+\frac{1}{2}\frac{b^2}{ml^3}.
\ee
Ziegler found that the domain of asymptotic stability for the damped pendulum is given by the inequalities
$P<P_k(b)$ and $b>0$ and he plotted it against the stability interval $P<P_k$ of the undamped system, Fig.~\ref{fig1}(b). Surprisingly, the limit of the critical load $P_k(b)$ when $b$ tends to zero turned out to be significantly lower than the critical load of the undamped system
\be{z7}
P_k^*=\lim_{b\rightarrow0}P_k(b)=\frac{41}{28}\frac{c}{l}\simeq1.464\frac{c}{l}<P_k.
\ee

Note that in the original work of Ziegler, formula \rf{z6} contains a misprint which yields linear dependency of the critical follower load on the damping coefficient $b$. Nevertheless, the domain of asymptotic stability found in \cite{Zi52} and reproduced in Fig.~\ref{fig1}(b), is correct.

\begin{figure}[h]
\includegraphics[width=0.8\textwidth]{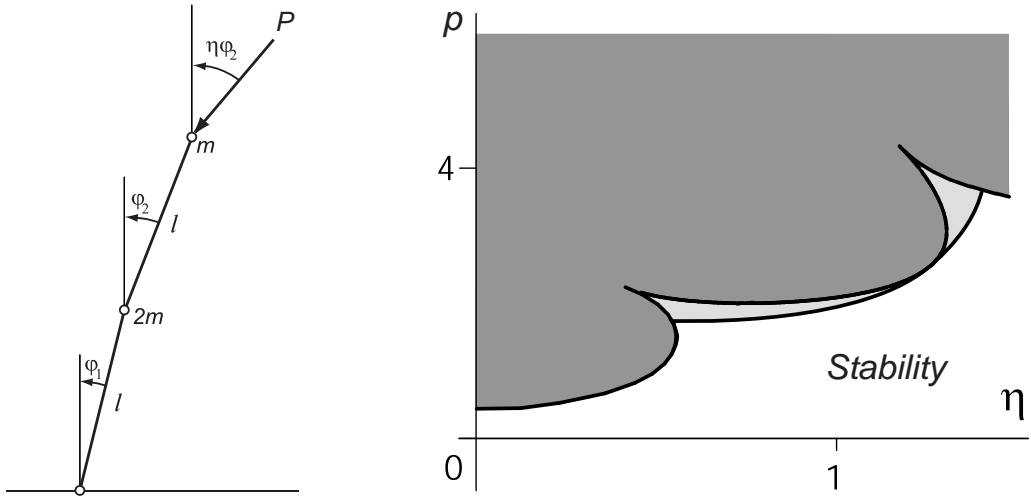}
\caption{\label{fig4} Ziegler's pendulum with the partially follower force \cite{K03a}: (dark gray) Instability domain in the absence of damping $(b_1=0, b_2=0)$ and (light gray) its increment in the presence of damping with $b_2 = 0.3 b_1$ and $b_1 \rightarrow 0$.}
\end{figure}

Some authors considered extensions of Ziegler's model by adding a conservative load and by assuming unequal
 damping coefficients \cite{BZ69,HJ65,K03a,Ko92,T95}. Fig.~\ref{fig4} demonstrates how the domain of instability for the undamped Ziegler's pendulum with the partially follower load ($\eta=1$ corresponds to the pure follower load), shown in dark gray in the $(\eta,p)$-plane, extends in a discontinuous manner in the presence of dissipation when $b_2=0.3 b_1$ and $b_1 \rightarrow 0$.
The portion of the stability domain that became unstable is depicted in light gray \cite{K03a,T95}. Therefore, the two-dimensional stability diagrams of the undamped system and the system with vanishingly small damping differ by a region of positive measure.

Ziegler drew attention both to the substantial decrease in the critical load of the damped non-conservative system with vanishingly small dissipation and to the high sensitivity of the critical follower load with respect to the variation of the damping distribution. In the mechanical engineering literature these two effects are called the Ziegler's paradox of destabilization by small damping.

In the conclusion to his classical book \cite{Bo63}, Bolotin emphasized that the discrepancy between the stability domains of undamped non-conservative systems and that of systems with infinitesimally small dissipation is a topic of the greatest theoretical interest in stability theory. Encouraging further research of the destabilization paradox, Bolotin was not aware that the crucial ideas for its explanation were formulated as early as 1956.

\section{Bottema's solution}
\label{Bot}

In 1956, in the journal `Indagationes Mathematicae', there appeared an article by Oene Bottema (1901-1992) \cite{B56},
then Rector Magnificus of the Technical University of Delft and an expert in classical geometry and mechanics,
that outstripped later findings for decades.
Bottema's work in 1955 \cite{B55} can be seen as an introduction, it was directly motivated by Ziegler's paradox. However, instead of examining the particular model of Ziegler, he studied in \cite{B56} a much more general class of
non-conservative systems.

Following \cite{B55,B56}, we consider a holonomic scleronomic linear system with two degrees of
freedom, of which the coordinates $x$ and $y$ are chosen in such a way that the kinetic energy is
$
T=({\dot x}^2+{\dot y}^2)/2.
$
Hence the Lagrange equations of small oscillations
near the equilibrium configuration $x = y = 0$ are as follows
\ba{b2}
\ddot x +a_{11}  x +a_{12}  y +b_{11} \dot x +b_{12} \dot y&=& 0,\nn \\
\ddot y +a_{21}  x +a_{22}  y +b_{21} \dot x +b_{22} \dot y&=& 0,
\ea
where $a_{ij}$ and $b_{ij}$ are constants, ${\bf A}:=(a_{ij})$ is the matrix of the forces depending on
the coordinates, ${\bf B}:=(b_{ij})$ of those depending on the velocities. If $\bf A$ is
symmetrical and while disregarding the damping associated with
the matrix $\bf B$, there exists a potential energy function
$V = (a_{11} x^2 + 2 a_{12}x y + a_{22} y^2)/2$,
if it is antisymmetrical, the forces are circulatory. When the matrix $\bf B$ is
symmetrical, we have a non-gyroscopic damping force, which is positive when
the dissipative function
$(b_{11} x^2 + 2 b_{12}x y + b_{22} y^2)/2$
is positive definite. If $\bf B$ is antisymmetrical the forces depending on the
velocities are purely gyroscopic.

The matrices $\bf A$ and $\bf B$ can both be written uniquely as the sum of symmetrical and
antisymmetrical parts: ${\bf A}={\bf K}+{\bf N}$ and ${\bf B}={\bf D}+{\bf G}$, where
\be{b3}
{\bf K}=\left(
          \begin{array}{cc}
            k_{11} & k_{12} \\
            k_{21} & k_{22} \\
          \end{array}
        \right), \quad {\bf N}=\left(
                  \begin{array}{rr}
                    0 & \nu \\
                    -\nu & 0 \\
                  \end{array}
                \right),\quad
                {\bf D}=\left(
          \begin{array}{cc}
            d_{11} & d_{12} \\
            d_{21} & d_{22} \\
          \end{array}
        \right), \quad {\bf G}=\left(
                  \begin{array}{rr}
                    0 & \Omega \\
                    -\Omega & 0 \\
                  \end{array}
                \right),
\ee
with $k_{11}=a_{11}$, $k_{22}=a_{22}$, $k_{12}=k_{21}=(a_{12}+a_{21})/2$, $\nu=(a_{12}-a_{21})/2$ and $d_{11}=b_{11}$, $d_{22}=b_{22}$, $d_{12}=d_{21}=(b_{12}+b_{21})/2$, $\Omega=(b_{12}-b_{21})/2$.

The system \rf{b3} has a potential energy function (disregarding damping) when $\nu= 0$, it is purely circulatory
for $k_{11} = k_{12}=k_{22} = 0$, it is non-gyroscopic for $\Omega= 0$, and has no damping when
$d_{11} = d_{12} = d_{22} = 0$. If damping exists, we suppose in this section that it is positive.

In order to solve the equations \rf{b3} we put $x = C_1 \exp(\lambda t)$, $y = C_2 \exp(\lambda t)$ and
obtain the characteristic equation for the frequencies of the small oscillations around equilibrium
\be{b5a}
Q:=\lambda^4 + a_1\lambda^3 + a_2\lambda^2 + a_3\lambda + a_4=0,
\ee
where \cite{K03a,K03b,K07a}
\be{b5}
a_1 = {\rm tr}{\bf D},\quad
a_2 = {\rm tr}{\bf K} + \det {\bf D} + \Omega^2,\quad
a_3 = {\rm tr}{\bf K}{\rm tr} {\bf D} - {\rm tr} {\bf KD} + 2\Omega\nu ,\quad
a_4 = \det{\bf K} + \nu^2.
\ee
For the equilibrium to be stable all roots of the characteristic equation \rf{b5a} must be semi-simple and have real parts which
are non positive.

It is always possible to write, in at least one way, the left hand-side as
the product of two quadratic forms with real coefficients,
$
Q=(\lambda^2+p_1\lambda+q_1)(\lambda^2+p_2\lambda+q_2).
$
Hence
\be{b7a}
a_1=p_1+p_2,\quad a_2=p_1p_2+q_1+q_2,\quad a_3=p_1q_2+p_2q_1,\quad a_4=q_1q_2.
\ee

For all the roots of the equation \rf{b5a} to be in the left side of the complex plane $(L)$ it is obviously necessary and sufficient
that $p_i$ and $q_i$ are positive. Therefore in view of \rf{b7a} we have: a necessary
condition for the roots $Q=0$ having negative real parts is $a_i>0$ $(i = 1, 2, 3,4)$.
This system of conditions however is not sufficient, as the example
$(\lambda^2-\lambda+2)(\lambda^2+2\lambda+3)=\lambda^4+\lambda^3+3\lambda^2+\lambda+6$ shows. But if $a_i>0$ it is not
possible that either one root of three roots lies in $L$ (for then $a_4\le0$); it
is also impossible that no root is in it (for, then $a_4 \le 0$). Hence if $a_i > 0$ at
least two roots are in $L$; the other ones are either both in $L$, or both on
the imaginary axis, or both in $R$. In order to distinguish between these cases we
deduce the condition for two roots being on the imaginary axis. If $\mu i$ ($\mu \ne 0$ is real)
is a root, then $\mu^4-a_2\mu^2+a_4=0$ and $-a_1\mu^2+a_3=0$.
Hence $H:=a_1^2a_4+a_3^2-a_1a_2a_3=0$.
Now by means of \rf{b7a} we have
\be{bb6}
H=-p_1p_2(a_1a_3+(q_1-q_2)^2).
\ee

In view of $a_1> 0$, $a_3>0$ the second factor is positive; furthermore
$a_1= p_1+p_2 > 0$, hence $p_1$ and $p_2$ cannot both be negative. Therefore $H < 0$
implies $p_1> 0$, $p_2> 0$, for $H = 0$ we have either $p_1 = 0$ or $p_2=0$ (and not both,
because $a_3>0$), for $H>0$ $p_1$ and $p_2$ have different signs. We see from
the decomposition of the polynomial \rf{b5a} that all its roots are in $L$ if $p_1$ and $p_2$ are positive.

Hence: a set of necessary and sufficient conditions for all roots of \rf{b5a}
to be on the left hand-side of the complex plane is
\be{bb6a}
a_i>0~~(i=1,2,3,4), \quad H<0.
\ee

We now proceed to the cases where all roots have non-positive
real parts, so that they lie either in $L$ or on the imaginary axis.
If three roots are in $L$ and one on the imaginary axis, this root must
be $\lambda = 0$. Reasoning along the same lines as before we find that necessary
and sufficient conditions for this are
$a_i>0~~(i=1,2,3)$, $a_4=0$, and $H<0$.
If two roots are in $L$ and two (different) roots on the imaginary axis we have
$p_1>0$, $q_1>0$, $p_2=0$, $q_2>0$ and the conditions are
$a_i>0~~(i=1,2,3,4)$ and $H=0$.
If one root is in $L$ and three are on the imaginary axis, then $p_1>0$, $q_1=0$, $p_2=0$,
$q_2>0$ and the conditions are
$a_i>0~~(i=1,2,3)$, $a_4=0$, and $H=0$.

The obtained conditions are border cases of \rf{bb6a}. This does not occur
with the last type we have to consider: all roots are on the imaginary axis. We
now have $p_1=0$, $p_2=0$, $q_1>0$, $q_2>0$. Hence $a_2 >0$, $a_4>0$, $a_1=a_3=0$ and
therefore $H = 0$. This set of relations is necessary, but not sufficient, as
the example $Q= \lambda^4 + 6\lambda^2+ 25 = 0$ (which has two roots in $L$ and two in the righthand side of the complex plane $(R)$)
shows. The proof given above is not valid because as seen from \rf{bb6a}, $H=0$
does not imply now $p_1p_2= 0$, the second factor being zero for $a_1a_3 = 0$
and $q_1 =q_2$. The condition can of course easily be given; the equation \rf{b5a} is
$\lambda^4+a_2\lambda^2+a_4=0$
and therefore it reads $a_2>0$, $a_4>0$, $a_2^2>4a_4$.

Summing up we have: all roots of \rf{b5a} (assumed to be different) have
non-positive real parts if and only if one of the two following sets of
conditions is satisfied \cite{B56}
\ba{bb11}
A:&~& a_1>0,~a_2>0,~a_3>0,~a_4\ge0,~a_2\ge \frac{a_1^2a_4+a_3^2}{a_1a_3},\nn \\
B:&~& a_1=0,~a_2>0,~a_3=0,~a_4>0,~a_2>2\sqrt{a_4}.
\ea

Note that $a_1$ represents the damping coefficients $b_{11}$ and $b_{22}$ in the system.
One could expect $B$ to be a limit of $A$, so that for $a_1 \rightarrow 0$, $a_3 \rightarrow 0$
the set $A$ would continuously tend to $B$. {\em That is not the case}.

\begin{figure}[h]
\includegraphics[width=0.9\textwidth]{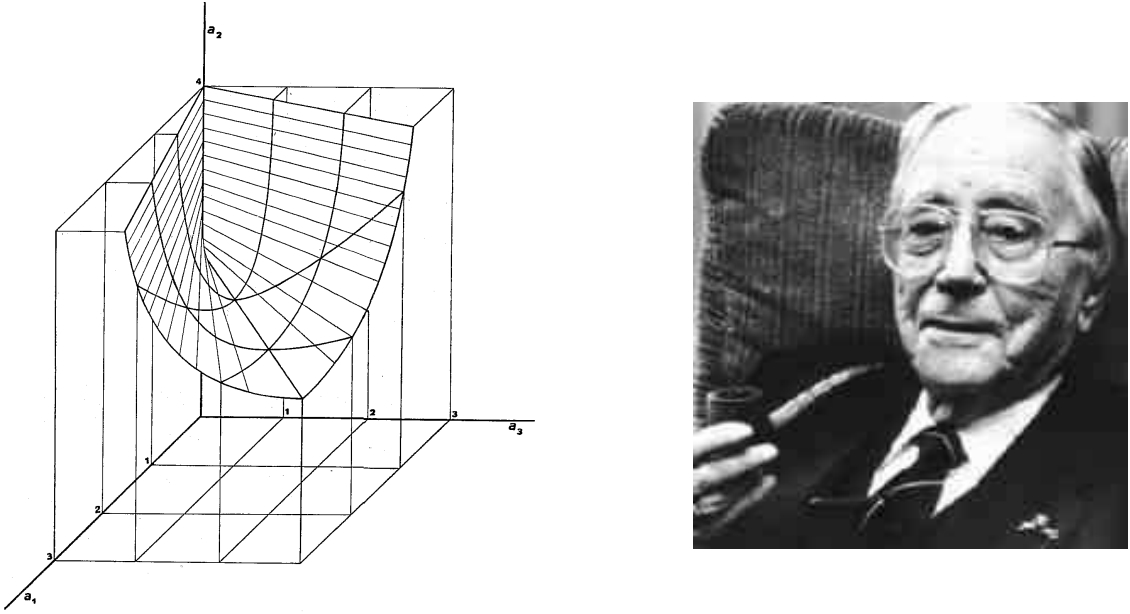}
\caption{\label{fig2} Original drawing (left) from the 1956 work \cite{B56} of Oene Bottema (right),
showing the domain of the asymptotic stability of the real polynomial of fourth order and of the two-dimensional non-conservative system
with Whitney's umbrella singularity. The ruled surface (called $V$ in the text) is given by equation
(\ref{bb7})}
\end{figure}

Remark first of all that the roots of \rf{b5a} never lie outside $R$ if $a_1 = 0$, $a_3 \ne 0$ (or $a_1 \ne  0$,
$a_3=0$). Furthermore, if $A$ is satisfied and we take $a_1=\varepsilon b_1$, $a_3=\varepsilon b_3$, where
$b_1$ and $b_3$ are fixed and $\varepsilon \rightarrow 0$, the last condition of $A$ reads $(\varepsilon \ne 0)$
$$
a_2 > \frac{b_1^2a_4+b_3^2}{b_1b_3}=g_1
$$
while for $\varepsilon = 0$ we have
$$
a_2> 2 \sqrt{a_4}=g_2.
$$
Obviously we have \cite{B56}
$$
g_1-g_2=\frac{(b_1 \sqrt{a_4}-b_3)^2}{b_1b_3}
$$
so that $(g_1>g_2)$ but for $b_3=b_1\sqrt{a_4}$. That means that in all cases where $b_3\ne b_1\sqrt{a_4}$ we have a discontinuity in our
stability condition. The phenomenon of the discontinuity was illustrated by Bottema in a geometrical diagram, Fig.~\ref{fig2}.

Following Bottema \cite{B56} we substitute in \rf{b5a} $\lambda=c\mu$, where $c$ is the positive fourth root
of $a_4>0$. The new equation reads $P:=\mu^4+ b_1\mu^3+ b_2\mu^2+b_3\mu+1 = 0$,
where $b_i=a_i/c^i$ $(i= 1,2,3,4)$. If we substitute $a_i=c^ib_i$ in $A$ and $B$ we get
the same condition as when we write $b_i$ for $a_i$, which was to be expected,
because if the roots of \rf{b5a} are outside $R$, those of $P=0$ are also outside $R$
and inversely. We can therefore restrict ourselves to the case $a_4 = 1$, so
that we have only three parameters $a_1$, $a_2$, $a_3$. We take them as coordinates
in an orthogonal coordinate system.

The condition $H=0$ or
\be{bb7}
a_1a_2a_3 = a_1^2 +a_3^2
\ee
is the equation of a surface $V$ of the third degree, which we have to
consider for $a_1 \ge 0$, $a_3 \ge 0$, Fig.~\ref{fig2}. Obviously $V$ is a ruled surface, the line $a_3=ma_1$,
$a_2=m+1/m$ $(0<m<\infty)$ being on $V$. The line is parallel to the $0a_1a_3$-plane
and intersects the $a_2$-axis in $a_1=a_3= 0$, $a_2=m+ 1/m\ge 2$. The $a_2$-axis
is the double line of $V$, $a_2>2$ being its active part.
Two generators pass through each point of it; they coincide for $a_2 = 2$ $(m = 1)$, and for $a_2\rightarrow\infty$
their directions tend to those of the $a_1$ and $a_3$-axis $(m=0, m=\infty)$.
The conditions $A$ and $B$ express that the image point $(a_1, a_2, a_3)$ lies
on $V$ or above $V$. The point $(0, 2, 0)$ is on $V$, but if we go to the $a_2$-axis
along the line $a_3=ma_1$ the coordinate $a_2$ has the limit $m+ 1/m$, which is
$>2$ but for $m= 1$. Curiously enough, even half a century later, there appear
papers repeating this reasoning and the result almost literally, see for instance \cite{S04}.

Note that we started off with 8 parameters in eq. \rf{b2}, but that the surface $V$ bounding the stability domain
is described by 3 parameters. It is described by a map of $E^2$ into $E^3$ as in Whitney's papers \cite{W43,W44}. Explicitly, a
transformation of (19) to (2) is given by
$$
a_1 =\frac{1}{2}y_3 + w,~~ a_2 = 2 + y_2,~~ a_3 = -\frac{1}{2}y_3 + w
$$
with $w^2 = \frac{1}{4}y_3^2+ y_1y_2$.

Returning to the non-conservative system \rf{b2} $(\nu \ne 0)$, with damping,
but without gyroscopic forces, so $\Omega = 0$, and assuming as in \cite{B55} that $k_{12}=0$, $k_{11}> 0$, and $k_{22} > 0$ (a similar setting but with $d_{12}=0$ and $k_{12}\ne0$ was considered later by Bolotin in \cite{Bo63,Bo02}), we find that the condition for stability $H\le0$ reads
\be{bb2}
\nu^2 < \frac{(k_{11}-k_{22})^2}{4}-\frac{(d_{11}-d_{22})^2(k_{11} - k_{22})^2-4(k_{11} d_{22} + k_{22} d_{11}) (d_{11}d_{22} - d_{12}^2) (d_{11} + d_{22})}{4(d_{11}+ d_{22})^2}.
\ee

Suppose now that the damping force decreases in a uniform way, so we put
 $d_{11} = \varepsilon d_{11}'$, $d_{12} = \varepsilon d_{12}'$, $d_{22} = \varepsilon d_{22}'$, where $d_{11}$, $d_{12}$, $d_{22}$ are constants and $\varepsilon \rightarrow 0$. Then, for the inequality \rf{bb2} we get
\be{bb3}
\nu^2 < \nu_{cr}^2:= \frac{(k_{11}-k_{22})^2}{4}-\frac{(d_{11}'-d_{22}')^2(k_{11} - k_{22})^2}{4(d_{11}'+ d_{22}')^2}.
\ee
But if there is no damping, we have to make use of condition $B$, which gives
\be{bb4}
\nu^2 < {\nu_0}^2:= \frac{(k_{11}-k_{22})^2}{4}=\left( \frac{{\rm tr}{\bf K}}{2}\right)^2-\det{\bf K}.
\ee
Obviously
\be{bb5}
{\nu_0}^2-\nu_{cr}^2=\frac{(d_{11}'-d_{22}')^2(k_{11} - k_{22})^2}{4(d_{11}'+ d_{22}')^2}=\left[\frac{2{\rm tr}{\bf K}{\bf D}-{\rm tr}{\bf K}{\rm tr}{\bf D}}{2{\rm tr}{\bf D}}\right]^2 \ge 0,
\ee
where the expressions written in terms of the invariants of the matrices involved \cite{K07a} are valid also without the restrictions on the matrices $\bf D$ and $\bf K$ that were adopted in \cite{Bo63,Bo02,B55}.
For the values of $\frac{2{\rm tr}{\bf K}{\bf D}-{\rm tr}{\bf K}{\rm tr}{\bf D}}{2{\rm tr}{\bf D}}$ which are small with respect to $\nu_0$ we can approximately write \cite{Ki04,K05}
\be{bb6}
\nu_{cr}\simeq\nu_0-\frac{1}{2\nu_0}\left[\frac{2{\rm tr}{\bf K}{\bf D}-{\rm tr}{\bf K}{\rm tr}{\bf D}}{2{\rm tr}{\bf D}}\right]^2.
\ee
If $\bf D$ depends on two parameters, say $\delta_{1}$ and $\delta_{2}$, then \rf{bb6} has a canonical form \rf{w} for the Whitney's umbrella in the $(\delta_1,\delta_2,\nu)$-space.
Due to discontinuity existing for  $2{\rm tr}{\bf K}{\bf D}-{\rm tr}{\bf K}{\rm tr}{\bf D}\ne0$ the equilibrium may be stable if there is no
damping, but unstable if there is damping, however small it may be. We observe also that the critical non-conservative parameter, $\nu_{cr}$, depends on the ratio of the damping coefficients and thus is strongly sensitive to the distribution of damping similarly to how it happens in rotor dynamics.
This is the results which Ziegler \cite{Zi52,Zi53} found in a special case.

\section{`Hopf meets Hamilton under Whitney's umbrella'}
\label{HmH}

The title of this section derives from a nice tutorial paper by Langford \cite{La03}.
As we have seen, Bottema was the first who established that the asymptotic stability domain of a real polynomial of fourth order in the space of its coefficients consists of one of the `pockets' of the Whitney umbrella.
The corresponding singularity was later identified as generic in the three parameter families of real matrices by V.I. Arnold
\cite{Ar71,Ar83}, who named it `deadlock of an edge'.
In this respect Bottema's results in \cite{B56} can be seen as an early study of bifurcations and structural stability of polynomials and matrices, and therefore of the singularities of their stability boundaries whose systematical
treatment was initiated since the beginning of the 1970s in \cite{Ar71,Ar83,L80,L82}.

Although Bottema applied his result to nonconservative systems without gyroscopic forces, there are
 reasons for the singularity  to appear in the case when gyroscopic forces are taken into account
because the stability is determined by the roots of a similar fourth order characteristic polynomial. In order to study this case
we consider separately the following $m$-dimensional version of the non-conservative system \rf{b2}
\be{i1}
\ddot {\bf x} + (\Omega{\bf G}+\delta{\bf D})\dot {\bf x} + ({\bf K}+\nu{\bf N}){\bf x} =0,
\ee
where a dot stands for time differentiation, ${\bf x}\in \mathbb{R}^m$, and real matrix
${\bf K}={\bf K}^T$ corresponds to potential forces. Real matrices ${\bf D}={\bf D}^T$,
${\bf G}=-{\bf G}^T$, and ${\bf N}=-{\bf N}^T$ are related to dissipative (damping), gyroscopic, and non-conservative
positional (circulatory) forces with magnitudes controlled by scaling factors
$\delta$, $\Omega$, and $\nu$ respectively.
A \textit{circulatory} system, to which the undamped Ziegler's pendulum is attributed \cite{K05,OR96,Se90}, is obtained from (\ref{i1}) by neglecting velocity-dependent forces
\be{i2}
\ddot {\bf x} + ({\bf K}+\nu{\bf N}){\bf x} =0,
\ee
while a \textit{gyroscopic} one has no damping and non-conservative positional forces
\be{i3}
\ddot {\bf x} + \Omega{\bf G}\dot {\bf x} + {\bf K}{\bf x} =0.
\ee
Circulatory and gyroscopic systems \rf{i2} and \rf{i3} possess fundamental symmetries that are evident
after transformation of equation \rf{i1} to the form $\dot{\bf y}={\bf C}{\bf y}$ with
\be{i4}
{\bf C}=
\left[%
\begin{array}{ll}
  -\frac{1}{2}\Omega{\bf G} & {\bf I} \\
  \frac{1}{2}\delta\Omega{\bf D}{\bf G}+\frac{1}{4}\Omega^2{\bf G}^2-{\bf K}-\nu{\bf N} &
  -\delta{\bf D}-\frac{1}{2}\Omega{\bf G} \\
\end{array}%
\right],~
{\bf y}=\left[%
\begin{array}{l}
  {\bf x} \\
  \dot{\bf x}{+}\frac{1}{2}\Omega{\bf G}{\bf x} \\
\end{array}%
\right],
\ee
where $\bf I$ is the identity matrix.

In the absence of damping and gyroscopic forces $(\delta=\Omega=0)$,
${\bf R}{\bf C}{\bf R}=-{\bf C}$ with
\be{i5}
{\bf R}={\bf R}^{-1}=\left[%
\begin{array}{cc}
  {\bf I} & 0 \\
  0 & -{\bf I} \\
\end{array}%
\right].
\ee
This means that the matrix $\bf C$ has a \textit{time reversal} symmetry, and equation \rf{i2}
describes a reversible dynamical system \cite{OR96}. Due to this property,
\be{i6}
\det({\bf C}-\lambda{\bf I})=\det({\bf R}({\bf C}-\lambda{\bf I}){\bf R})=\det({\bf C}+\lambda{\bf I}),
\ee
and the eigenvalues of circulatory system \rf{i2} appear in pairs $(-\lambda, \lambda)$.
Without damping and non-conservative positional forces $(\delta=\nu=0)$ the matrix $\bf C$
possesses the \textit{Hamiltonian} symmetry ${\bf J}{\bf C}{\bf J}={\bf C}^T$,
where $\bf J$ is a symplectic matrix \cite{Ar83, MK91, Br94} with
\be{i7}
\quad {\bf J}=-{\bf J}^{-1}=\left[%
\begin{array}{cc}
  0 & {\bf I} \\
  -{\bf I} & 0 \\
\end{array}%
\right].
\ee
As a consequence,
\be{i8}
\det({\bf C}-\lambda{\bf I})=\det({\bf J}({\bf C}-\lambda{\bf I}){\bf J})=\det({\bf C}^T+\lambda{\bf I})=
\det({\bf C}+\lambda{\bf I}),
\ee
which implies that if $\lambda$ is an eigenvalue of $\bf C$ then so is $-\lambda$, similar
to the reversible case.
Therefore, an equilibrium of a circulatory or of a gyroscopic system
is either unstable or all its eigenvalues lie
on the imaginary axis of the complex plane, in the last case implying marginal stability if they are semi-simple.

It is well known that in the Hamiltonian case, the transition from gyroscopic stability to flutter instability occurs through the interaction of simple purely imaginary eigenvalues with the opposite Krein signature known as the Krein collision or the Hamiltonian Hopf bifurcation \cite{Ha92,La03,MK86,MK91,MO95}. The collision occurs at the border of marginal stability, say at $\Omega=\Omega_0$ for \rf{i3}, and it yields a double pure imaginary eigenvalue with the Jordan chain of vectors, which splits into a a complex conjugate pair under destabilizing variation of the parameter $\Omega$.

Let $i\omega_0$ be the double eigenvalue at $\Omega=\Omega_0$ with the Jordan chain of generalized eigenvectors ${\bf u}_0$, ${\bf u}_1$, satisfying the equations \cite{K07b}
\ba{h14}
(-{\bf I}\omega_0^2+i\omega_0\Omega_0{\bf G}+{\bf K}){\bf u}_0&=&0,\nn \\
(-{\bf I}\omega_0^2+i\omega_0\Omega_0{\bf G}+{\bf K}){\bf u}_1&=&
-(2i\omega_0{\bf I}+\Omega_0{\bf G}){\bf u}_0.
\ea
Then, the Krein collision in the gyroscopic system \rf{i3} is described by the following expressions
\ba{h19}
i\omega(\Omega)&=&i\omega_0\pm i\mu\sqrt{\Omega-\Omega_0}+o(|\Omega-\Omega_0|^{\frac{1}{2}}), \nn \\
{\bf u}(\Omega)&=&{\bf u}_0\pm i\mu{\bf u}_1\sqrt{\Omega-\Omega_0}+o(|\Omega-\Omega_0|^{\frac{1}{2}}),
\ea
where the real coefficient $\mu$ is according to  \cite{K07b}
\be{h20}
\mu^2=-\frac{2\omega_0^2\overline{\bf u}_0^T{\bf u}_0}
{\Omega_0^+(\omega_0^2\overline{\bf u}_1^T{\bf u}_1-\overline{\bf u}_1^T{\bf K}{\bf u}_1-
i\omega_0\Omega_0^+\overline{\bf u}_1^T{\bf G}{\bf u}_1-\overline{\bf u}_0^T{\bf u}_0)}
\ee
with the bar over a symbol denoting complex conjugate.

Perturbing the system \rf{i3} by small damping and circulatory forces yields an increment to a simple pure imaginary eigenvalue \cite{K05,K07b}
\be{h18}
\lambda=i\omega(\Omega)-\frac{\omega^2(\Omega) \overline{\bf u}^T(\Omega) {\bf D}{\bf u}(\Omega)\delta-
i\omega\overline{\bf u}^T(\Omega){\bf N}{\bf u}(\Omega)\nu}{\overline{\bf
u}^T(\Omega) {\bf K} {\bf u}(\Omega)+\omega^2 \overline{\bf u}^T(\Omega) {\bf u}(\Omega)}+o(\delta,\nu).
\ee

With the expressions \rf{h19}, equation \rf{h18} is used for the calculation of the deviation from the imaginary axis of the eigenvalues that participated in the Krein collision in the presence of the non-Hamiltonian perturbation that makes the merging of modes an {\em imperfect} one \cite{HG03}.

Since $\bf D$ and $\bf K$ are real symmetric matrices and $\bf N$ is a real skew-symmetric one,
the first-order increment to the eigenvalue $i\omega(\Omega)$ given by \rf{h18} is real-valued.
Consequently,
in the first approximation in  $\delta$ and $\nu$, the
simple  eigenvalue $i\omega(\Omega)$ remains on the imaginary axis, if
$\nu=\gamma(\Omega)\delta$, where
\be{h21}
\gamma(\Omega)=-i\omega(\Omega)\frac{\overline{\bf
u}^T(\Omega) {\bf D}{\bf u}(\Omega)}{\overline{\bf u}^T(\Omega){\bf N}{\bf
u}(\Omega)}.
\ee
With the expansions \rf{h19} the formula \rf{h21} reads
\be{h22}
\gamma(\Omega)=-(\omega_0\pm \mu\sqrt{\Omega-\Omega_0})\frac{d_1\mp \mu d_2\sqrt{\Omega-\Omega_0}}
{n_1\pm \mu n_2\sqrt{\Omega-\Omega_0}},
\ee
where we define
\ba{h15}
d_1&=&{\rm Re}(\overline{\bf u}_0^T{\bf D}{\bf u}_0),\quad
d_2={\rm Im}(\overline{\bf u}^T_0 {\bf D}{\bf u}_1-\overline{\bf u}^T_1 {\bf D}{\bf u}_0),\nn \\
n_1&=&{\rm Im}(\overline{\bf u}_0^T{\bf N}{\bf u}_0),\quad
n_2={\rm Re}(\overline{\bf u}^T_0 {\bf N}{\bf u}_1-\overline{\bf u}^T_1 {\bf N}{\bf u}_0),\quad
\gamma_*=-i\omega_0\frac{\overline{\bf u}_0^T {\bf D}{\bf
u}_0}{\overline{\bf u}_0^T{\bf N}{\bf u}_0}.
\ea
From \rf{h22} it follows that in the vicinity of $\gamma:=\nu/\delta=\gamma_*$ the limit of the critical value of the gyroscopic parameter
$\Omega_{cr}$ of the near-Hamiltonian system as $\delta \rightarrow 0$ exceeds the threshold of gyroscopic stabilization
determined by the Krein collision (see \cite{K07b})
\be{h17}
\Omega_{cr}(\gamma)=\Omega_0 +\frac{n_1^2 (\gamma-\gamma_*)^2}{\mu^2 (\omega_0 d_2 -\gamma_* n_2 - d_1)^2}\ge\Omega_0.
\ee

Substituting $\gamma=\nu\delta$ in expression \rf{h17} yields a simple estimate
for the critical value of the gyroscopic parameter $\Omega_{cr}(\delta,\nu)$ that has a canonical form \rf{w}
and therefore describes the Whitney's umbrella surface in the $(\delta,\nu,\Omega)$-space \cite{K07b}
\be{h23}
\Omega_{cr}(\delta,\nu)=\Omega_0+\frac{n_1^2 (\nu-\gamma_*\delta)^2}
{\mu^2 (\omega_0 d_2 -\gamma_* n_2 - d_1)^2\delta^2}.
\ee
In case of two oscillators ($m=2$) the approximation \rf{h23} is transformed to \cite{K06,K07a,K07b}
\be{h13}
\Omega_{cr}(\delta,\nu)=\Omega_0 +
\Omega_0\frac{2}{(\omega_0{\rm tr}{\bf D})^2\delta^2}(\nu - \gamma_*\delta)^2, \quad
\gamma_{*}:=\frac{{\rm tr}{\bf K}{\bf D}+({\Omega_0}^2-\omega_0^2){\rm tr}{\bf D}}{2\Omega_0},
\ee
where $\omega_0=\sqrt[4]{\det{\bf K}}$ and $\Omega_0=\sqrt{-{\rm tr}{\bf K}+2\sqrt{\det{\bf K}}}$ in the assumption that $\det{\bf K}>0$ and ${\rm tr}{\bf K}<0$.
Due to the singularity the gyroscopic stabilization in the presence of dissipative and non-conservative positional
forces depends on the ratio $\frac{\nu}{\delta}$ and is thus very sensitive to non-Hamiltonian perturbations.
We will discuss gyroscopic stabilization in more detail in section \ref{mbe}.

We note that the sensitivity of simple eigenvalues of Hamiltonian and gyroscopic systems to dissipative perturbations was
a subject of intensive investigations, see, e.g., MacKay \cite{MK91}, Haller \cite{Ha92}, and Bloch et al. \cite{Br94}.
MacKay pointed out the necessity to extend such a perturbation analysis to multiple eigenvalues \cite{MK91}.
Maddocks and Overton \cite{MO95} initiated the study of multiple eigenvalues and showed that for an appropriate class of dissipatively perturbed Hamiltonian systems, the number of unstable modes of the dynamics linearized at a nondegenerate equilibrium is determined solely by the index of the equilibrium regarded as a critical point of the Hamiltonian.
They analyzed the movement of the eigenvalues in the limit of vanishing dissipation without direct application, however, to
the destabilization paradox and approximation of the singular stability boundary. Our calculations performed in this section use the ideas developed in \cite{K06,K07a,K07b,K09} that, however, can be traced back to the works of Andreichikov and Yudovich \cite{AY75} and Crandall \cite{C95}.

We see that in Hamiltonian mechanics, the Hamiltonian-Hopf bifurcation in which two pairs of complex conjugate eigenvalues approach the imaginary axis symmetrically from the left and right, then merge in double purely imaginary eigenvalues and separate along the imaginary axis (or the reverse) has codimension one. In the general case of non-Hamiltonian vector fields, the occurrence of double imaginary eigenvalues has codimension three.
The interface between these two cases possesses the Whitney umbrella singularity; the Hamiltonian
systems lie on its handle. Quoting Langford from his introductory paper [55] linking Hopf bifurcation, Hamiltonian
mechanics and Whitneys umbrella: `Hopf meets Hamilton
under Whitney's umbrella', which, we add, was  opened by Bottema.

\section{Parametric resonance in systems with dissipation.}
\label{Par}
Parametric resonance arises usually in applications if we have an independent (periodic) source
of energy. The classical example is the mathematical pendulum with oscillating support and a typical
equation studied in this context is the Mathieu equation:
\[ \ddot{x} + (\omega^2 + \varepsilon \cos \nu t)x=0. \]
In the case of this equation, basic questions are: for what values of the parameters
$\omega, \varepsilon, \nu$ is the trivial solution $x= \dot{x}=0$ stable or unstable? Another
basic question is, what happens on adding damping effects? In the theory, certain resonance relations
between the frequencies $\omega$ and $\nu$ play a crucial part.
See for instance \cite{Ar83}, \cite{Bo63}, \cite{Sey}, \cite{YS75} or \cite{V09} and Fig. \ref{Mat}(a) for
this classical case.

In applications with parametric excitation where usually more degrees of freedom play a
part, many combination
resonances are possible. For a number of interesting cases, analysis and more references see
\cite{Bo63,Sey}. In what follows, the so-called sum resonance will be important.

First we will consider the general procedure for systems with this combination resonance,
after which we will discuss an application.

\subsection{Normalization of oscillators in sum resonance}
\label{sum}

In \cite{HR95} a geometrical explanation is presented for damping induced instability in parametric
systems using `all' the parameters of the system as unfolding
parameters. It will turn out that, using symmetry and normalization, four parameters are needed
to give a complete
description in a two degrees of freedom system, or more generally systems where three
frequencies are in resonance, but three parameters suffice to visualize the situation.\\
Consider the following type of nonlinear differential equation with three frequencies
\begin{equation}
\label{vgl.devgl}
\dot{\bf x} = {\bf A}{\bf x} + \varepsilon {\bf f}({\bf x}, \omega_0 t),\;{\bf x} \in {\mathbb{R}}^4,
\end{equation}
which describes for instance a system of two parametrically forced coupled oscillators.
$\bf A$ is a $4 \times 4$ matrix, containing a number of parameters, with purely imaginary eigenvalues
$\pm i\omega_1$ and $\pm i\omega_2$. Assume that $\bf A$ is semi-simple, so, if necessary,
we can put $\bf A$ into diagonal form.
The vector valued function $\bf f$  contains both linear and nonlinear terms and is
$2\pi$-periodic in $\omega_0 t$, ${\bf f}(0, \omega_0 t)=0$ for all $t$. Eq. (\ref{vgl.devgl})
can be resonant in many different
ways, but as announced, we consider here the sum resonance
\[ \omega_1+\omega_2=\omega_0, \]
where the system may exhibit instability.
The parameter $\delta$ is used to control the
detuning $\delta = (\delta_1, \delta_2)$ of the frequencies $(\omega_1, \omega_2)$
near resonance and the parameter  $\mu = (\mu_1, \mu_2)$ derives from the damping coefficients.
So we may put ${\bf A}={\bf A}(\delta, \mu)$.
We summarize the analysis from \cite{HR95}.

\begin{figure}[h]
\begin{center}
\resizebox{10cm}{!}{
\includegraphics{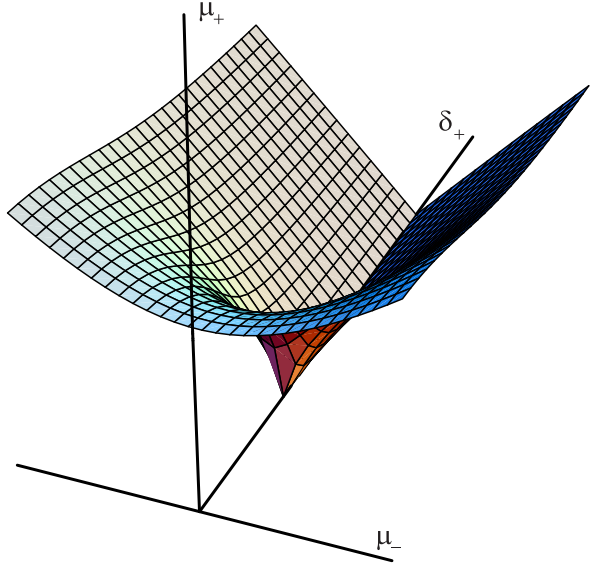}}
\end{center}
\caption{The critical surface in $(\mu_+,\mu_-,\delta_+)$ space for eq. (\ref{vgl.devgl}).
$\mu_+=\mu_1+\mu_2$, $\mu_-=\mu_1-\mu_2$, $\delta_+=\delta_1+\delta_2$. Only
the part $\mu_+>0$ and $\delta_+>0$ is shown. The parameters $\delta_1, \delta_2$
control the detuning of the frequencies, the parameters $\mu_1, \mu_2$ the damping
of the oscillators (vertical direction).
The base of the umbrella lies along the $\delta_{+}$-axis.
\label{umbrel}}
\end{figure}

The basic approach will be to put eq. (\ref{vgl.devgl}) into normal form by
normalization or averaging whereas the theory from \cite{Ar71} will play a part.  In
the normalized equation the
time-dependence is removed from lower order and appears only in the higher order terms.
It turns out that the autonomous, linear part
of this equation contains already enough information to determine the stability regions
of small amplitude oscillations near the origin. The linear part of
the normal form can be written as
\[ \dot{\bf z}={\bf A}(\delta, \mu){\bf z} \]
with 4-dimensional
\begin{equation}
\label{vgl.nf}
{\bf A}(\delta,\mu)=
\left(\begin{array}{cc}
{\bf B}(\delta,\mu) & 0\\
0 & \overline{\bf B}(\delta,\mu)
\end{array}\right),
\end{equation}
and
\begin{equation}
{\bf B}(\delta,\mu)=
\left(\begin{array}{cc}
i\delta_1 - \mu_1 & \alpha_1 \\
\overline{\alpha}_2 & -i\delta_2 - \mu_2
\end{array}\right).
\end{equation}

 Since ${\bf A}(\delta,\mu)$ is the complexification of a real matrix, it commutes with
complex conjugation. Furthermore, according to the normalization described in
\cite{Ar83}, \cite{IA92} and \cite{SVM} and
if $\omega_1$ and $\omega_2$ are independent over the
integers, the normal form of eq. (\ref{vgl.devgl}) has a continuous symmetry
group.
The second step is then to test the linear part ${\bf A}(\delta,\mu)$ of the normalized
equation for structural stability i.e. to answer the question whether there exist
open sets in parameter space where the dynamics is qualitatively the same. The analysis
follows \cite{Ar71} and \cite{Ar83}.
The family of matrices ${\bf A}(\delta,\mu)$ is parameterized by the detuning
$\delta$  and the damping  $\mu$. The procedure is to
 identify the most degenerate member $\bf N$ of this family, which turns out to be
${\bf A}(\delta, 0)$ and then show that
${\bf A}(\delta,\mu)$ is its versal unfolding in the sense of \cite{Ar83}.
The family ${\bf A}(\delta,\mu)$ is equivalent to a versal unfolding
of the degenerate member ${\bf N}$. For details we refer again to \cite{HR95,V09},
an explicit example is discussed in the next subsection.

We can put the conclusions in a different way:
 the family ${\bf A}(\delta,\mu)$ is structurally stable for $\delta, \mu >0$, whereas
${\bf A}(\delta, 0)$ is not. This has interesting consequences in applications as  small damping
and zero damping may exhibit very different behavior.
 In parameter space, the stability regions of the trivial solution
are separated by a {\em critical surface} which is the hypersurface where
${\bf A}(\delta,\mu)$ has at least one pair of purely imaginary complex conjugate
eigenvalues. As before, this critical surface is diffeomorphic to the {\em Whitney
umbrella}, see Fig. \ref{umbrel}. It is the singularity of the Whitney
umbrella that causes the discontinuous behavior displayed in the stability
diagram in the subsection \ref{rotor2}. The structural stability argument guarantees that the results are
`universally valid', i.e. they qualitatively hold for generic systems in sum resonance.

Above we have described the basic normalization approach, but if we are interested only
in the shape of the resonance (instability) tongues, there are faster methods. For instance
using the Poincar\'e-Linstedt method, see \cite{V09}.

\subsection{Rotor dynamics without damping}
\label{rotor1}
The effects of adding linear damping to a parametrically excited system
have already been observed and described in for instance \cite{Bo63}, \cite{YS75}, \cite{SS90}, or \cite{Sey}.
The following example is based on  \cite{RTV93}.

\begin{figure}[h]
\begin{center}
\resizebox{6cm}{!}{
\includegraphics{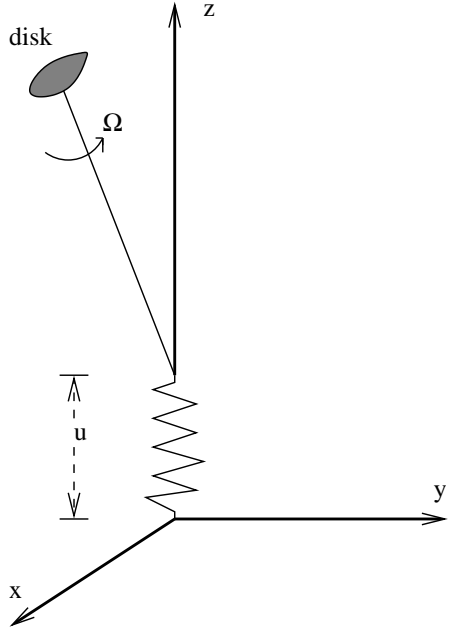}}
\end{center}
\caption{Rotor with diskmass $M$, elastically mounted with axial ($u$)
and lateral directions.
 \label{rotation}}
\end{figure}

Consider a rigid rotor consisting of a heavy disk
of mass $M$ which is rotating with constant rotation speed $\Omega$ around an axis.
The axis of rotation is elastically
mounted on a foundation; the connections which are holding the
 rotor in an upright position are also elastic. To describe the position of the rotor we
have the axial displacement $u$ in the vertical direction (positive upwards), the angle of the axis of
 rotation { \it with respect to} the $z$-axis and {\it around} the $z$-axis.
Instead of these two angles we will use the projection of the center of gravity motion on the
horizontal $(x, y)$-plane, see Fig. \ref{rotation}.
Assuming small oscillations in the upright ($u$) position, frequency $2 \eta$, the equations
of motion without damping become after rescaling:
\begin{eqnarray}
\label{Msys}
\ddot{x} + 2 \alpha \dot{y} + ( 1 + 4 \varepsilon \eta^2 \cos 2 \eta t)x &=& 0,\nonumber \\
\ddot{y} - 2 \alpha \dot{x} + ( 1+ 4 \varepsilon \eta^2 \cos 2 \eta t) y &=& 0.
\end{eqnarray}\noindent
The parameter $\alpha$ is proportional to the rotation speed $\Omega$.
System (\ref{Msys}) constitutes a conservative system of coupled Mathieu-like equations.
 Abbreviating $P(t)=  4  \eta^2 \cos 2 \eta t$,
the corresponding Hamiltonian is:
\[ H= \frac{1}{2}(1+ \alpha^2 + \varepsilon P(t))x^2 + \frac{1}{2}p_x^2 +
\frac{1}{2}(1+ \alpha^2 + \varepsilon P(t))y^2 + \frac{1}{2}p_y^2 + \alpha xp_y - \alpha yp_x, \]
where $p_x, p_y$ are the momenta. The natural frequencies of
the unperturbed system (\ref{Msys}), $ \varepsilon = 0,$  are $\omega_1 = \sqrt{ \alpha^2 + 1}
+ \alpha $
and $\omega_2 = \sqrt{ \alpha^2 +1 } - \alpha.$
By putting $z = x + iy$, system (\ref{Msys}) can be written as:
\begin{equation}
\ddot{z} - 2 \alpha i \dot{z} + (1 + 4 \varepsilon \eta^2 \cos 2 \eta t)z = 0. \ \ \
\end{equation}\noindent
Introducing the new variable:
\begin{equation}
v = e^{ -i \alpha t} z, \ \ \
\end{equation}\noindent
and rescaling time $\eta t = \tau $, we obtain:
\begin{equation}
\label{veq}
v^{\prime \prime} + \left( \frac{ 1 + \alpha^2 }{ \eta^2 } + 4 \varepsilon  \cos 2
\tau \right) v = 0,
\end{equation}\noindent
where the prime denotes differentiation with respect to $\tau .$
By writing down the real and imaginary parts of this equation, we have actually got
two identical Mathieu equations.

Using the classical and well-known results on the Mathieu equation,
we conclude that the trivial solution is stable for $\varepsilon $ small enough, provided that
$\sqrt{ 1 + \alpha^2 }$ is not close to $n \eta$ , for $n=1,2,3,...$.
The first-order and most prominent interval of instability, $n=1,$ arises if:
\begin{equation}
\label{unst}
\sqrt{1 + \alpha^2 } \approx \eta.
\end{equation}\noindent
If condition (\ref{unst}) is satisfied, the trivial solution of equation (\ref{veq}) is unstable.
Therefore, the trivial solution of system (\ref{Msys}) is also unstable.
 Note that this instability arises when:
\[ \omega_1 + \omega_2 = 2\eta,\]
i.e. when the sum of the eigenfrequencies of the unperturbed system equals the
excitation frequency $2 \eta $ which is the  sum resonance of first order.
The domain of instability is bounded by:
\begin{equation}
\label{bound}
 \eta_b = \sqrt{ 1 + \alpha^2 }~(1 \pm \varepsilon ) + O( \varepsilon^2 ) \ \ \ .
\end{equation}
See Fig. \ref{Mat}(b) where the V-shaped instability domain is presented in the case of
rotor rotation ($\alpha \neq 0$) without damping.

Higher order combination resonances can be studied in the same way; the domains of
instability in parameter space continue to narrow as $n$ increases. As noted,
 the parameter $\alpha$ is proportional to the rotation speed $\Omega$ of the disk and
also to the ratio of the moments of inertia.

\subsection{Rotor dynamics with damping}
\label{rotor2}

We add small linear damping to system (\ref{Msys}), with positive damping parameter
$\mu = 2 \varepsilon \kappa $. This leads to the equations:
\begin{eqnarray}
\label{Msysd}
\ddot{x} + 2 \alpha \dot{y} + ( 1 + 4 \varepsilon \eta^2 \cos 2 \eta t)x +2 \varepsilon \kappa \dot{x} &=& 0,\nonumber \\
\ddot{y} - 2 \alpha \dot{x} + ( 1+ 4 \varepsilon \eta^2 \cos 2 \eta t) y+2 \varepsilon \kappa \dot{y} &=& 0.
\end{eqnarray}\noindent
and using the complex variable $z$:
\begin{equation}
\label{zdamp}
\ddot{z} - 2\alpha i \dot{z} + \left( 1 + 4\varepsilon \eta^2 \cos 2\eta t \right) z +
 2 \varepsilon \kappa \dot{z} = 0.
\end{equation}\noindent

Because of the damping term, we can no longer reduce the complex eq. (\ref{zdamp}) to two
identical second order real equations, as we did previously.

In the sum resonance of the first order, we have $\omega_1+\omega_2 \approx 2 \eta$
and the solution of the unperturbed $( \varepsilon = 0)$ equation can be written as:
\begin{equation}
z(t) = z_1 e^{i \omega_1 t} + z_2 e^{-i\omega_2 t}, ~~~~z_1, z_2 \in {\mathbb{C}},
\end{equation}
with $\omega_1 = \sqrt{ \alpha^2 +1} + \alpha ,~~~\omega_2 = \sqrt{ \alpha^2 + 1}
 - \alpha.$

Applying variation of constants leads to equations for $z_1$ and $z_2$:
\begin{eqnarray}
\label{zdamp1}
\dot{z}_1 &=&\frac{i \varepsilon }{\omega_1 + \omega_2 } ( 2\kappa (i \omega_1 z_1 - i
\omega_2 z_2 e^{-i(\omega_1 + \omega_2 )t} ) +
\nonumber \\
& & \ \ \ 4\eta^2 \cos 2 \eta t (z_1 + z_2 e^{-i(\omega_1 + \omega_2 )t} ) ),\nonumber \\
  \dot{z}_2 &=& \frac{ -i \varepsilon }{\omega_1 + \omega_2 } ( 2\kappa (i\omega_1 z_1
 e^{i( \omega_1 + \omega_2 )t} - i\omega_2 z_2 ) +
\nonumber \\
& & \ \ \ 4\eta^2 \cos 2 \eta t (z_1 e^{i( \omega_1 + \omega_2 )t} + z_2 ) ).
\end{eqnarray}

\begin{figure}
\begin{center}
\includegraphics[width=0.9\textwidth]{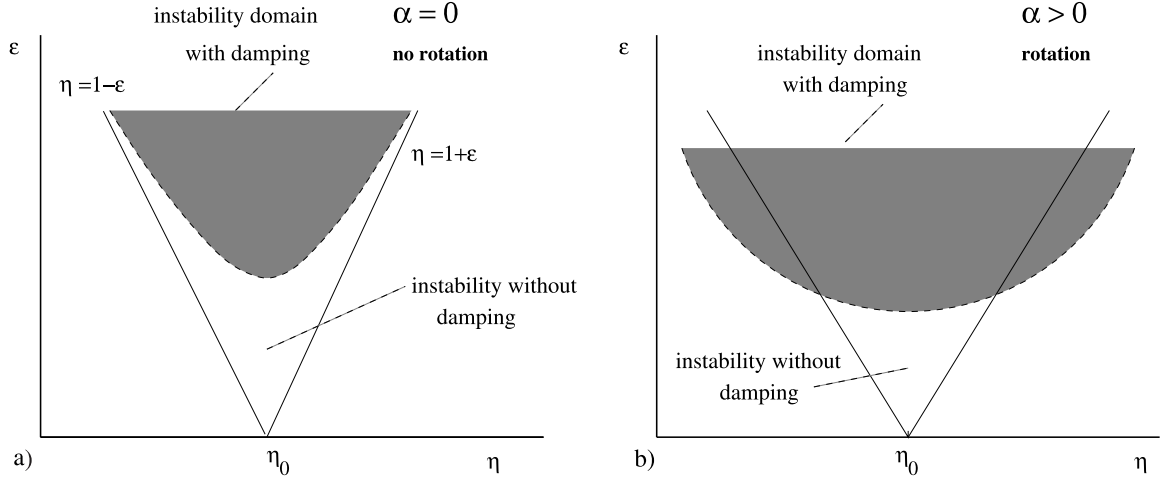}
\end{center}
\caption{\label{Mat} (a) The classical case as we find for instance for the Mathieu equation
with and without damping; in the case of damping the instability tongue is lifted off from the
$\eta$-axis and the instability domain is reduced. (b) The instability tongues
for the rotor system. Again, because of damping the instability tongue is lifted off from the $\eta$-axis,
but the tongue broadens. The boundaries of the $V$-shaped tongue without damping
are to first approximation described by
the expression $\eta = \sqrt{1 + \alpha^2}(1 \pm \varepsilon)$, $\eta_0 =  \sqrt{1 + \alpha^2}$.}
\end{figure}

To calculate the instability interval around the value
$\eta_0 = \frac{1}{2} (\omega_1 + \omega_2 ) = \sqrt{ \alpha^2 + 1}$,
we apply normal form or (periodic solution) perturbation theory, see \cite{RTV93} for details,
to find for the stability boundary:
\begin{eqnarray}
\label{bound2}
\eta_b & = &  \sqrt{ 1 + \alpha^2 } \left( 1 \pm \varepsilon \sqrt{ 1 + \alpha^2 -
\frac{ \kappa^2 }{ \eta_{0}^{2} }} + ....\right) \ \ \ ,\nonumber \\
 &=& \sqrt{ 1 + \alpha^2 }
\left( 1 \pm \sqrt{ (1+ \alpha^2 ) \varepsilon^2 - \left( \frac{\mu }{2 \eta_0 } \right)^2 }
+ .... \right) \ \ \ .
\end{eqnarray}

It follows  that, as in other examples we have seen, the domain of instability actually
becomes {\em larger} when damping is introduced. See Fig. \ref{Mat}b.

The instability interval,
shows a discontinuity at $\kappa=0$.

If $\kappa  \rightarrow 0$, then the boundaries of the instability domain
tend to the limits $\eta_b \rightarrow \sqrt{ 1 + \alpha^2 } ( 1 \pm \varepsilon
\sqrt{1 + \alpha^2 } )$ which differs from the result we found when $\kappa = 0:~\eta_b =
\sqrt{1 + \alpha^2 } (1 \pm \varepsilon ).$ For reasons of comparison, we display the instability
tongues in Fig. \ref{Mat} in the four cases with and without rotation, with and without damping.

Mathematically, the bifurcational behavior is again described by the Whitney umbrella as indicated in
subsection \ref{sum}.
In mechanical terms, the broadening of the instability-domain is caused by the
coupling between the two degrees of freedom of the rotor in lateral directions which arises in
the presence of damping.

\section{Manifestation of the destabilization paradox in other applications}
\label{app}

In this section we discuss additional applications from physics and engineering, both finite- and infinite-dimensional.
\subsection{Gyroscopic systems of rotor dynamics}
\label{mbe}

Investigation of the stability of equilibria of the Hauger's \cite{Ha75} and Crandall's \cite{C95,Sa08}
gyropendulums as well as
of the Tippe Top \cite{B08,KM07} and the Rising Egg \cite{B08} leads to the system of linear equations known as
the modified Maxwell-Bloch equations \cite{Br94}.

The modified Maxwell-Bloch equations are the normal form
for rotationally symmetric, planar dynamical systems \cite{Br94, B08}.
They follow from the equation \rf{i1} for $m=2$, ${\bf D}={\bf I}$, and ${\bf K}=\kappa{\bf I}$, where $\kappa$ corresponds to potential forces,
and thus can be written as a single differential equation with complex coefficients
\be{e1}
\ddot{x}+i\Omega\dot{x}+\delta\dot{x}+i\nu{x}+\kappa{x}=0,~~ x=x_1-ix_2.
\ee
According to \rf{bb6a} the solution $x=0$ of equation \rf{e1} is asymptotically stable if
and only if
\be{e2}
\delta > 0 ,\quad
\Omega> \frac{\nu}{\delta}-\frac{\delta}{\nu} \kappa.
\ee

\begin{figure}[h]
\includegraphics[width=0.99\textwidth]{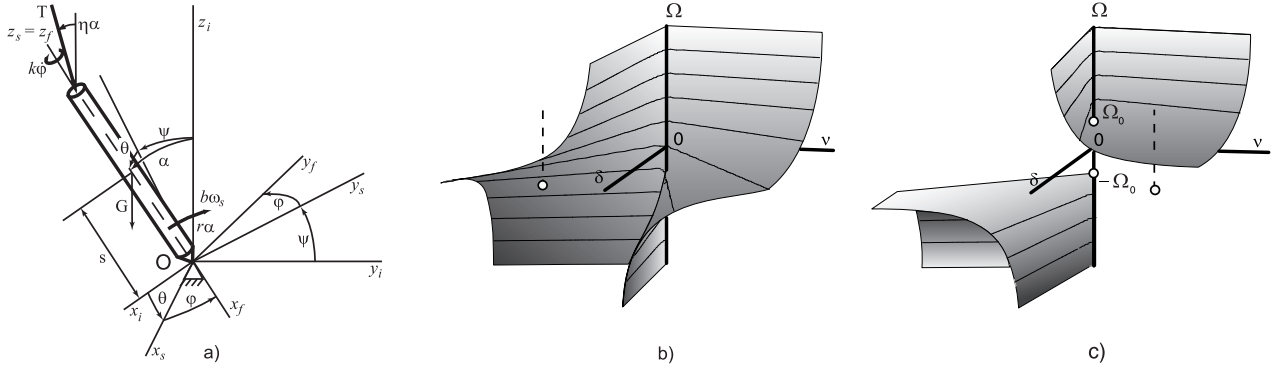}
\caption{\label{fig5} (a) Hauger's gyropendulum; (b) Dissipation-induced destabilization of its statically stable equilibrium $(\kappa>0)$ in the presence of circulatory forces; (c) Singular domain of gyroscopic stabilization of its statically unstable equilibrium $(\kappa<0)$ in the presence of non-Hamiltonian perturbations yields the destabilization paradox.}
\end{figure}

For $\kappa>0$ the domain of asymptotic stability is a dihedral angle with the $\Omega$-axis serving
as its edge, Fig.~\ref{fig5}(b). Its sections by the planes $\Omega=const$
are contained in the angle-shaped regions with the boundaries
\be{e3}
\nu=\frac{\Omega\pm\sqrt{\Omega^2+4\kappa}}{2}\delta.
\ee
At $\Omega=0$ the angle is
bounded by the lines $\nu=\pm\delta\sqrt{\kappa}$ and thus it is less than $\pi$.
The domain of asymptotic stability is twisting around the $\Omega$-axis in such a manner
that it always remains in the half-space $\delta>0$, Fig.~\ref{fig5}(b). Consequently,
the system that is statically stable at $\Omega=0$ and $\delta \ge 0$ can become unstable at greater $\Omega$ in the presence
of non-conservative positional forces,
as shown in Fig.~\ref{fig5}(b) by the dashed line.
The larger magnitudes of circulatory forces, the lower $|\Omega|$ at the onset of instability.
This is a typical example of {\em dissipation-induced instability} in the sense of
\cite{Br94,KM06,KM07,KM09} when
only non-Hamiltonian perturbations can cause the destabilizing movements of eigenvalues with definite Krein signature \cite{K08}.

As $\kappa>0$ decreases, the hypersurfaces forming the dihedral angle
approach each other so that, at $\kappa=0$, they temporarily merge along the line $\nu=0$
and a new configuration originates for $\kappa<0$, Fig.~\ref{fig5}(c). The new
domain of asymptotic stability consists of two disjoint parts
that are pockets of two Whitney's umbrellas singled out by inequality $\delta>0$.
The absolute values of the
gyroscopic parameter $\Omega$ in the stability domain are always not less than
$\Omega_0=2\sqrt{-\kappa}$. As a consequence, the system that is statically unstable at $\Omega=0$ can become
asymptotically stable at greater $\Omega$ in the presence of circulatory forces,
as shown in Fig.~\ref{fig5}(c) by the dashed line.

As a mechanical example we consider Hauger's gyropendulum \cite{Ha75}, which is an axisymmetric
rigid body of mass $m$ hinged at the point $O$ on the axis of symmetry as shown in Fig.~\rf{fig5}(a).
The body's moment of inertia with respect to the axis through the point $O$
perpendicular to the axis of symmetry is denoted by $I$, the body's moment of inertia
with respect to the axis of symmetry is denoted by $I_0$, and the distance between the fastening point
and the center of mass is $s$. The orientation of the pendulum, which is associated with the trihedron
$Ox_fy_fz_f$, with respect to the fixed trihedron $Ox_iy_iz_i$ is specified by the angles $\psi$, $\theta$, and
$\phi$. The pendulum experiences the force of gravity $G = mg$ and a follower torque $T$
that lies in the plane of the $z_i$ and $z_f$ coordinate axes. The moment vector makes an angle of $\eta\alpha$
with the axis $z_i$, where $\eta$ is a parameter ($\eta \ne 1$) and $\alpha$ is the angle between the $z_i$
and $z_f$ axes. Additionally, the pendulum experiences the restoring elastic moment $R = -r\alpha$ in the
hinge and the dissipative moments $B = -b\omega_s$ and $K = -k\phi$, where $\omega_s$ is the angular velocity
of an auxiliary coordinate system $Ox_sy_sz_s$ with respect to the inertial system and $r$, $b$, and $k$
are the corresponding coefficients.

Linearization of the nonlinear equations of motion derived in \cite{Ha75} with the new variables $x_1=\psi$ and
$x_2=\theta$ and the subsequent nondimensionalization yield the Maxwell-Bloch equations \rf{e1}
where the dimensionless parameters are given by
\be{e4}
\Omega=\frac{I_0}{I},~\delta=\frac{b}{I\omega},~\kappa=\frac{r-mgs}{I\omega^2},~\nu=\frac{1-\eta}{I\omega^2}T,~~
\omega=-\frac{T}{k}.
\ee
The domain of asymptotic stability of the Hauger gyropendulum, given by \rf{e2}, is shown in Fig.~\ref{fig5}(b,c).

For the statically unstable gyropendulum $(\kappa<0)$
the singular points on the $\Omega$-axis correspond to the critical values $\pm \Omega_0 = \pm 2 \sqrt{-\kappa}$
and the critical frequency $\omega_0 = \sqrt{-\kappa}$.
We find approximations of the stability boundary near the Whitney umbrella
singularity as derived in \cite{K07b,K09}:
\be{e6}
\Omega_{cr}(\nu,\delta)=
\pm2\sqrt{-\kappa}\pm\frac{1}{\sqrt{-\kappa}}\frac{(\nu\mp\delta\sqrt{-\kappa})^2}{\delta^2}.
\ee
Thus, Hauger's gyropendulum, which is statically unstable at $\Omega = 0$, can become asymptotically stable for sufficiently
large $|\Omega| \ge \Omega_0$ under a suitable distribution of dissipative and nonconservative positional forces. For almost all
combinations of $\delta$ and $\nu$ the onset of gyroscopic stabilization of the non-conservative system is greater than that of a pure gyroscopic one (destabilization paradox: $\Omega_{cr}(\nu,\delta)\ge \Omega_0$). The obtained results are valid also for the equilibria of Tippe Top, Rising Egg, and Crandall's gyropendulum \cite{K06,K07a}.

\subsection{Circulatory systems of rotor dynamics}

In some rotor dynamics applications gyroscopic effects are neglected \cite{K39,KM07}. For example,
in the modeling of friction-induced oscillations in disc- and drum brakes, clutches and other machinery,
the speed of rotation is assumed to be small. This frequently yields the linearized equations of motion in the form of a circulatory system with or without damping. In recent models the damping is included because it is believed that high sensitivity of the
squeal onset to the damping distribution might be responsible for the poor reproducibility of the laboratory experiments
with the squealing machinery.

Hoffmann and Gaul \cite{HG03} studied a model of a mass sliding over a conveyor belt with friction and detected that
small damping in this circulatory system destroys the reversible Hopf bifurcation and makes the collision of eigenvalues imperfect,
exactly as it happens with the eigenvalues of Ziegler's pendulum \cite{K05,K07c}.

\begin{figure}
\includegraphics[width=0.9\textwidth]{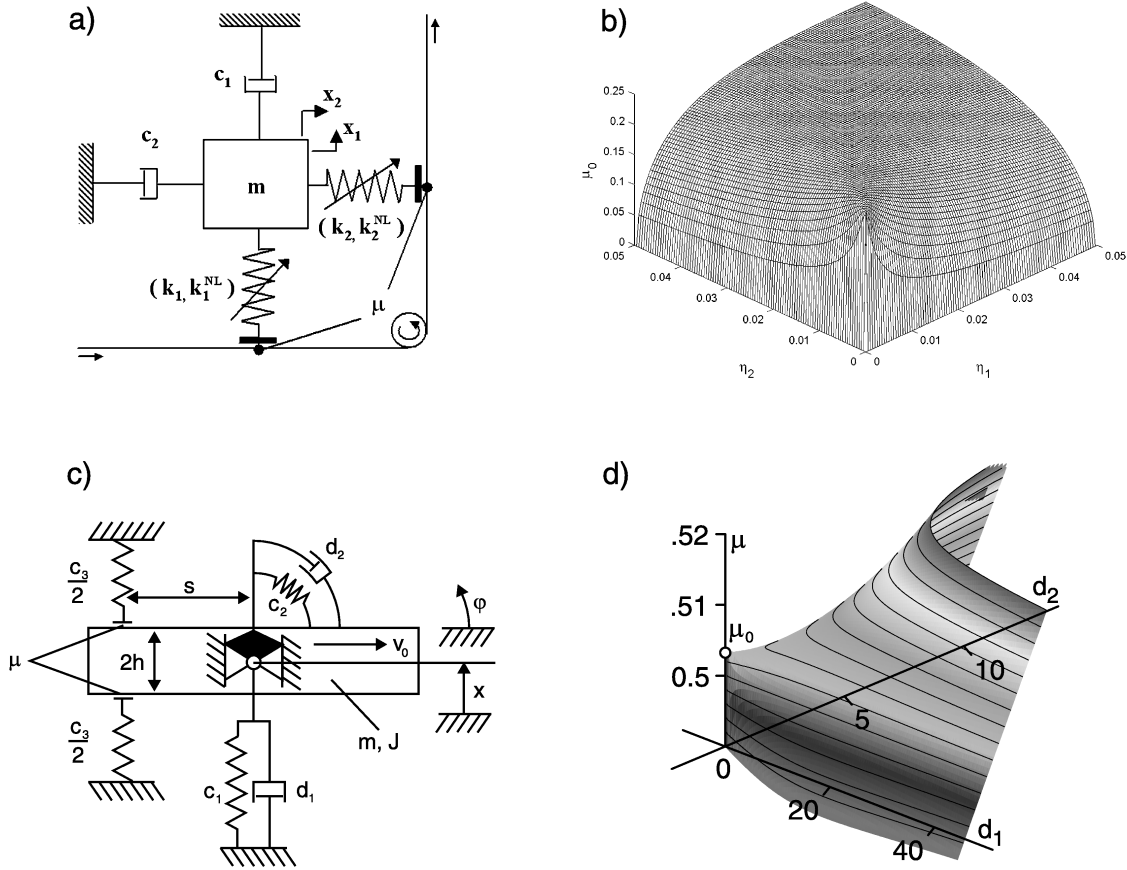}
\caption{\label{fig6} (a) A model of a drum brake \cite{H93}; (b) Its numerically calculated critical friction coefficient at the onset of flutter instability as a function of damping parameters \cite{SJ07}; (c) A model of a disc brake \cite{PP02}; (d) Its critical friction coefficient at the onset of flutter instability as a function of damping parameters calculated both numerically and perturbatively \cite{K07c}.}
\end{figure}

In order to study squeal vibration in drum brakes ${\rm Hult\acute{e}n}$ \cite{H93,SJ07} introduced a model shown in Fig.~\ref{fig6}(a).
This model is composed of a mass $m$ held against a moving band; the contact between the mass and the
band is modeled by two plates supported by two different springs. It is assumed that the
mass and band surfaces are always in contact and that the
contact can be expressed by two cubic stiffnesses. Damping is included as shown in Fig.~\ref{fig6}(a). The friction coefficient
at contact is assumed to be constant and the band moves at a constant velocity. Then it is assumed that the direction
of friction force does not change because the relative velocity between the band speed and $\dot x_1$ or $\dot x_2$ is assumed to
be positive.
The tangential force $F_T$ due to friction contact is assumed to be proportional to the normal force $F_N$ as given by
Coulomb's law: $F_T = \mu F_N$. Assuming the normal force $F_N$ is linearly related to the displacement of the mass normal
to the contact surface, the resulting equations of motion can be expressed as
$$
\left(
  \begin{array}{cc}
    1 & 0 \\
    0 & 1 \\
  \end{array}
\right) \ddot{\bf x}+\left(
                       \begin{array}{cc}
                         \eta_1\omega_{0,1} & 0 \\
                         0 & \eta_2\omega_{0,2} \\
                       \end{array}
                     \right)\dot{\bf x}+\left(
                               \begin{array}{cc}
                                 \omega_{0,1}^2 & -\mu \omega_{0,2}^2 \\
                                 \mu \omega_{0,1}^2 & \omega_{0,2}^2 \\
                               \end{array}
                             \right){\bf x}=0,
$$
being exactly of the form considered by Bottema. Here the relative damping coefficients are denoted by $ \eta_i = c_i/\sqrt{m_ik_i}$ $ (i = 1, 2)$ and natural pulsations are $\omega_{0,i} =\sqrt{k_i/m_i}$ $(i = 1, 2)$. Fig.~\ref{fig6}(b) shows the numerically calculated domain of asymptotic stability of the drum brake in the space of the friction coefficient $\mu$ and two damping coefficients $\eta_1$ and $\eta_2$ with the Whitney umbrella singularity \cite{SJ07}.

In Fig.~\ref{fig6}(c) a model of a disc brake proposed in \cite{PP02} is demonstrated. Its linearized equations of motion are again
that of a circulatory system with small damping. It is not surprising that the critical friction coefficient at the onset of
friction-induced vibrations as a function of two damping coefficients is represented in Fig.~\ref{fig6}(d) by a surface with the Whitney umbrella singularity \cite{K07c}.

In both examples a selected distribution of damping exists that yields an increase in the critical load rather than decrease that happens for all other distributions.
This possibility for stabilization was first pointed out in \cite{S96} for the Ziegler's pendulum. We will discuss this effect below in more detail.

\subsection{Infinite-dimensional near-reversible and near-Hamiltonian systems}
Dynamic instability, or flutter, is a general phenomenon
which commonly occurs in coupled fluid-structure
systems including pipes conveying fluids and  airfoils  \cite{Bo63,H67,HD92}.
Typically, the models are finite dimensional or continuous reversible systems that
demonstrate the destabilization paradox in the presence of damping. In a recent study
\cite{Z07} Ziegler's paradox was observed in a problem of a vocal fold vibration (phonation) onset.

\subsubsection{Near-reversible case: Beck's column with external and internal damping
}

Beck's column loaded by a follower force is a paradigmatic model for studying dynamical instability of structures.
In 1969 Bolotin and Zhinzher \cite{BZ69} investigated the effects of damping distribution on its stability.
They considered on the interval $x\in[0,1]$ the non-selfadjoint boundary eigenvalue problem of the form
\be{be1}
Lu:=N(q)u+\lambda D(d_1,d_2) u + \lambda^2 M u =0,\quad  {\bf U}{\bf u}:={\bf U}_N(q){\bf u} + \lambda {\bf U}_D (d_1,d_2){\bf u} +\lambda^2{\bf U}_M {\bf u} = 0,
\ee
where $\lambda$ is an eigenvalue with the eigenfunction $u(x)$.
The operators in the differential expression
\be{be2}
N=\partial^4_x +q \partial^2_x,\quad D = d_1 \partial^4_x+d_2 I, \quad M = I
\ee
depend on the magnitude of the follower load $q$ and the parameters of external, $d_2$, and internal (Kelvin-Voight), $d_1$, damping.
The matrices of boundary conditions in \cite{BZ69} are ${\bf U}_D=0$, ${\bf U}_M =0$, and
\be{be3}
{\bf U}_N=\left(
               \begin{array}{cccccccc}
                 1 & 0 & 0 & 0 & 0 & 0 & 0 & 0 \\
                 0 & 1 & 0 & 0 & 0 & 0 & 0 & 0 \\
                 0 & 0 & 0 & 0 & 0 & 0 & 1 & 0 \\
                 0 & 0 & 0 & 0 & 0 & 0 & 0 & 1 \\
               \end{array}
             \right),
\ee
and the vector ${\bf u}=(u(0),\partial_x u(0),\partial^2_x u(0),\partial^3_x u(0),u(1),\partial_x u(1),\partial^2_x u(1),\partial^3_x u(1))^T$. Some authors considered different boundary conditions that depend both on the physical parameters and
on the spectral parameter \cite{PS87,Z94}

The undamped Beck's column is stable for $q<q_0\simeq20.05$ \cite{CM79}. Stability is lost at $q\ge q_0$ when after the reversible Hopf bifurcation
the double pure imaginary eigenvalue $i\omega_0\simeq11.02$ splits into a pair of complex eigenvalues.
In \cite{BZ69} it was found that in the presence of infinitesimally small Kelvin-Voight damping the critical load is reduced
to $q=q_{cr}\simeq10.94$ and the critical frequency drops to $\omega=\omega_{cr}\simeq 5.40$.

There were numerous attempts to find an approximation of the new critical load by studying
the splitting of the double eigenvalue $i\omega_0$ of the unperturbed reversible system due
to dissipative perturbations \cite{Se90}.
Banichuk et al. \cite{BBM89a,BBM89} have emphasized the importance of degenerate perturbations,
the linear part of which
is in the tangent plane to the Whitney umbrella singularity.
Nevertheless, their analysis is not complete.

Further development of the approach of \cite{BBM89a,BBM89} in \cite{K03a,K05,KS05a,KS05b,KS05c,SK03}
resulted in the approximation to the critical load in the form
\be{be4}
q_{\rm cr}({\bf d})= q_{0}+ \frac{(\langle {\bf f}, {\bf d}
\rangle{+}\langle {\bf H} {\bf d}, {\bf d} \rangle)^2}
{\widetilde{f}\langle{\bf h}, {\bf d} \rangle^2}-
\frac{\omega_0^2}{\widetilde{f}} \langle {\bf G} {\bf d}, {\bf
d}\rangle,
\ee
where the vector of the damping parameters ${\bf {d}}{=}({d}_1, {d}_{2})$ and angular brackets denote the
scalar product in $\mathbb{R}^{2}$. The components of the vector ${\bf f}$ and the real scalar $\widetilde{f}$ are
\be{be5}
f_{r}=\left(\frac{\partial {D}}{\partial d_r}u_0, v_0\right)+
{\bf v}_0^*\widetilde{\bf V}_0^*\frac{\partial {\bf U}_D}{\partial d_r}{\bf u}_0,
~~
\widetilde{f}{=}\left(\frac{\partial {N}}{\partial q}u_0, v_0\right)+
{\bf v}_0^*\widetilde{\bf V}_0^*\frac{\partial {\bf U}_N}{\partial q}{\bf u}_0,~~
r=1, 2,
\ee
and the components of the vector ${\bf h}$  are defined as
\be{be6}
i{h}_r=
\left(\frac{\partial {D}}{\partial d_r}u_1, v_0\right)+
\left(\frac{\partial {D}}{\partial d_r}u_0, v_1\right)+
{\bf v}_1^*\widetilde{\bf V}_0^*\frac{\partial {\bf U}_D}{\partial d_r}{\bf u}_0+
{\bf v}_0^*\widetilde{\bf V}_0^*\frac{\partial {\bf U}_D}{\partial d_r}{\bf u}_1+
{\bf v}_0^*\left(\frac{\partial \widetilde{\bf V}}{\partial \bar{\lambda}}\right)^*\!\!
\frac{\partial {\bf U}_D}{\partial d_r}{\bf u}_0,~~
r=1, 2
\ee
with the asterisk denoting complex conjugate transposition and $(u,v)=\int_0^1 u(x)\bar v(x) dx$. The derivatives are taken at ${\bf d}=0$ and $q=q_0$ corresponding to the eigenvalue $\lambda=i\omega_0$ with the eigen- and associated functions $u_0$ and $u_1$.
The real matrix $\bf H$ has the components
\be{be7}
H_{r\sigma}=\frac{1}{2}\left( \frac{\partial^2 {D}}{\partial
d_r \partial d_{\sigma}} u_0, v_0\right)+
\frac{1}{2}{\bf v}_0^*\widetilde{\bf V}_0^*\frac{\partial^2 {\bf U}_D}{\partial d_r\partial d_{\sigma}}{\bf u}_0,
~~ r,\sigma=1,2
\ee
and the real matrix ${\bf G}$ is defined by the expression
\be{be8}
\langle {\bf G} {\bf d},{\bf d} \rangle= \sum_{r=1}^{2}
 d_r \left( \frac{\partial {D}}{\partial d_r}\hat{w}_2, v_0+{\bf v}_0^*\widetilde{\bf V}_0^*\frac{\partial {\bf U}_D}{\partial d_r}\hat{\bf w}_2\right),
\ee
where $\hat{w}_2$ is the solution of the boundary value problem
\be{be9}
N(q_0)\hat{w}_2-\omega_0^2M\hat{w}_2=
\sum_{r=1}^{2}{d}_{r}\frac{\partial D}{\partial d_{r}}u_0,
\quad
{\bf U}_N(q_0)\hat{\bf w}_2-\omega_0^2{\bf U}_M\hat{\bf w}_2=\sum_{r=1}^{2}{d}_{r}\frac{\partial {\bf U}_D}{\partial d_{r}}{\bf u}_0.
\ee
The eigenfunctions $u_0$ and $v_0$ and the associated functions $u_1$ and $v_1$ of the original and adjoint eigenvalue problems
are chosen to satisfy the bi-orthogonality and normalization conditions
\ba{be10}
2i\omega_0(M u_1, v_1)&+&(M u_0, v_1)+(M u_1, v_0)+ (\tilde{\bf V}_0{\bf v}_1 +\tilde{\bf V}'_{\bar \lambda}{\bf v}_0)^* (2i\omega_0{\bf U}_M{\bf u}_1 +{\bf U}_M{\bf u}_0)
+ {\bf v}_0^*\tilde{\bf V}_0^*{\bf U}_M {\bf u}_1 = 0,\nn\\
2i\omega_0(Mu_1, v_0)&+& 2i\omega_0 {\bf v}_0^*\tilde{\bf V}_0^*{\bf U}_M {\bf u}_1=1,
\ea
where the adjoint boundary value problems are connected by the Lagrange formula
\be{be11}
(Lu ,v) - (u, L^*v)= ({\bf V}{\bf v})^*\tilde{\bf U}{\bf u}- (\tilde{\bf V}{\bf v})^*{\bf U}{\bf u}.
\ee

\begin{figure}[h]
\includegraphics[width=0.99\textwidth]{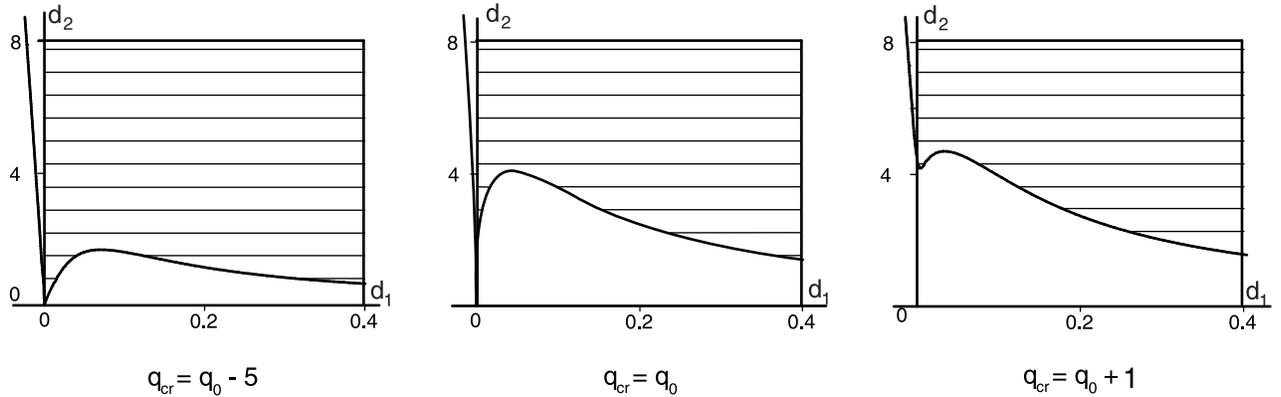}
\caption{\label{fig7} (Hatched) Cross sections of the approximation \rf{be12} to the stability domain of the damped Beck's column showing the possibility of the increase of the critical load by small damping.}
\end{figure}

Formula \rf{be4} can serve for the approximation of the jump in the critical load. In the
finite dimensional case with two degrees of freedom the expression for the limit of the critical
load reduces to \rf{bb6} \cite{K05}. For the Beck column described by the equations \rf{be1}
we calculate the critical load as \cite{KS05a,KS05b}
\be{be12}
q_{\rm cr}(d_1, d_2)=
q_0-\frac{1902d_1^2}{(14.34d_1+0.091d_2)^2}+
12.68d_1d_2+0.053d_2^2.
\ee
Additionally, \rf{be4} captures more information on the geometry of the stability domain. For example one can plot the cross sections of the stability domain \rf{be12} for the different levels of $q$ and find that some combinations of internal and external damping {\em increase} the critical load for the Beck's column, as shown in the central and right pictures of Fig.~\ref{fig7}.
The form of the stability boundary with the Whitney umbrella singularity approximated by
equation \rf{be12} was confirmed later by numerical computations in \cite{KOS07}. The limit
in the critical load following from \rf{be12} agrees well with the numerical data of \cite{AY75}.

Structural mechanics also has examples of near-Hamiltonian continuous systems showing discontinuous
changes in the stability domain.
As a modern application we mention a moving beam with frictional contact investigated in \cite{S08}.
Below we will consider an interesting example of the occurrence of the destabilization paradox in fluid dynamics.

\subsubsection{Near-Hamiltonian case: The instability of baroclinic zonal currents}

In the 1940s the first studies appeared of instability of baroclinic zonal (west-east) currents in the Earth's
 atmosphere \cite{CE49,E49}. It is remarkable that the unexpected destabilizing effect due to
the introduction of dissipation was
discovered in the linear stability analyses of this hydrodynamical problem by Holopainen (1961)
\cite{Ho61} and Romea (1977) \cite{R77} at the very same period of active research on the
destabilization paradox in structural mechanics. Recently these studies
were revisited by Krechetnikov and Marsden \cite{KM09} with the aim to handle rigorously the
treatment of dissipation-induced instability.

Romea considered an infinite channel in the periodic zonal direction $x$ of width $L$ in
the meridional direction $y$ that is rotating with an angular velocity $\Omega$.
Two layers of incompressible,
homogeneous fluids of slightly different densities (the
lighter fluid on top) are confined by the side walls and
by horizontal planes, a distance $D$ apart. For simplicity,
it is assumed that, in the absence of motion, the interface
is located halfway between the horizontal planes, and is
flat so that centrifugal effects may be ignored. Each
layer moves downstream with a constant velocity and
the slope of the interface is related to these velocities
through the thermal wind relation. It is implicitly
assumed that this basic state is maintained against
dissipation by an external energy source which is
unimportant with respect to the rest of the problem \cite{R77}.

The linearized equations for each layer near the basic state, characterized by
 the geostrophic streamfunctions $-U_1 y$ and $-U_2 y$, are according to \cite{R77,KM09}:
\ba{bar1}
(\partial_t + U_1 \partial_x)[\nabla^2\varphi_1+F (\varphi_2-\varphi_1)]+[\beta+F(U_1-U_2)]\partial_x\varphi_1&=&-r\nabla^2\varphi_1,\nn\\
(\partial_t + U_2 \partial_x)[\nabla^2\varphi_2+F (\varphi_1-\varphi_2)]+[\beta-F(U_1-U_2)]\partial_x\varphi_2&=&-r\nabla^2\varphi_2.
\ea
where $F$ is the internal rotational Froude number, $r\ge0$ is the measure of the effect of Ekman suction (Ekman layer dissipation), and $\beta$ is the planetary vorticity factor introduced to take into account
the variation of the Coriolis parameter with latitude ($\beta$-effect).

Assuming the wave solutions $\varphi_{1,2}\sim e^{i\alpha(x-ct)}\sin(m\pi y)$, where real $\alpha > 0$ is the $x$ wavenumber, Romea obtained a dispersion relation for the complex phase speed $c=c_r+i c_i$ in the form of the second-order complex polynomial. The real part of $c$ is the speed of propagation of the perturbation, while $\alpha c_i$ is the growth rate of the wave. If $c_i>0$, the wave grows, and the system is unstable.

\begin{figure}[h]
\includegraphics[width=0.95\textwidth]{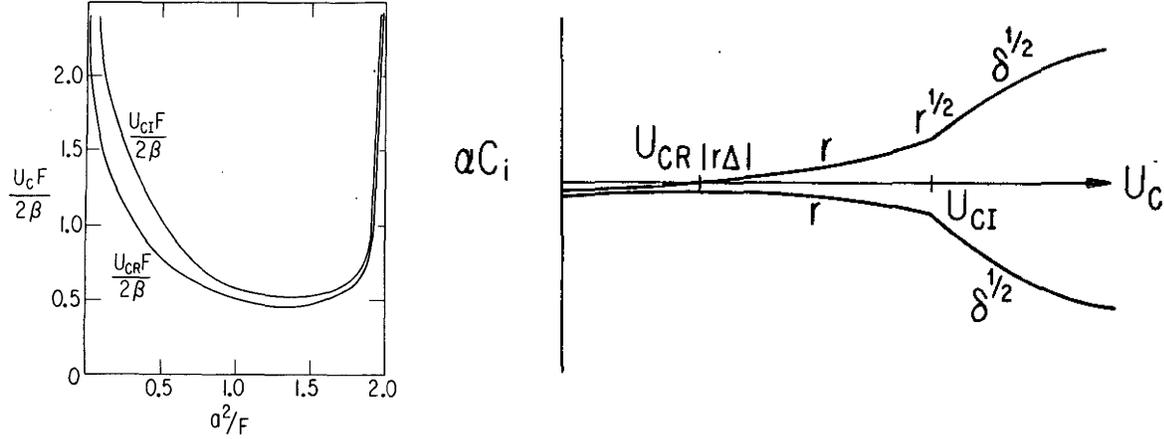}
\caption{\label{fig8} Original drawings from the 1977 work of Romea \cite{R77}: (left) Critical shear as a function of wavenumber demonstrates a discontinuous transition from the case when the Ekman layer dissipation $r=0$ initially $(U_{cI})$ to the case when $r\rightarrow 0$ $(U_{cR})$; (right) a typical imperfect merging of modes (growth rates) that substitutes the `perfect' Krein collision in near-Hamiltonian and near-reversible systems and is characteristic for the destabilization paradox \cite{BZ69}, \cite{Bo02}.}
\end{figure}

In the inviscid case when the Ekman layer dissipation is set to zero, the transition to instability occurs through the Krein collision that occurs at $U_c:=U_1-U_2=U_{cI}$, where \cite{R77,KM09}
\be{bar2}
U_{cI}=\frac{2\beta F}{a^2\sqrt{4F^2-a^4}}
\ee
with $a^2=\alpha^2+m^2\pi^2$. The critical shear $U_{cI}$ as a function of the wavenumber is plotted in Fig.~\ref{fig8}(left). This curve
bounds the region of marginal stability of the system without dissipation.

In the limit of vanishing viscosity $(r\rightarrow 0)$, the stability boundary differs from \rf{bar2}
\be{bar3}
U_{cR}=\frac{2\beta F}{a(a^2+F)\sqrt{2F-a^2}}.
\ee
The discrepancy between the stability domains of viscous and inviscid systems is clearly seen in Fig.~\ref{fig8}(left).
Therefore, Romea demonstrated that an introduction of infinitesimally small dissipation destabilizes the system,
lowering the curve of marginal stability by an $O(1)$ amount. This is the appearance of the destabilization paradox in a continuous near-Hamiltonian system, which is similar to that found in near-reversible systems like Ziegler's pendulum (cf. Fig.~\ref{fig4}) and Beck's column with dissipation \cite{BZ69,T95}. Fig.~\ref{fig8}(right) reproduces the original drawing from \cite{R77} showing the typical imperfect merging of modes \cite{HG03} that substitutes the `perfect' Krein collision in near-Hamiltonian and near-reversible systems. Approximation to the eigenvalue branches in imperfect merging can be efficiently calculated by means of the  perturbation theory of multiple eigenvalues for a wide class of non-conservative systems \cite{K03a,Ki04,K05,KS05b,KS05c,K08p}.

\section{Conclusion}

We have revisited the pioneering result of Oene Bottema who in 1956 resolved the paradox of
destabilization
by small damping and interpreted it by means of what is now called the Whitney umbrella singularity. We have
shown that this phenomenon
frequently occurs in near-Hamiltonian and near-reversible systems originating in very different areas
of mechanics and physics
ranging from hydrodynamics to contact mechanics and we have presented a unified treatment of it.
There are a few related topics upon we did not touch.
We mention interesting connections of this effect
to structured pseudospectra \cite{KZ07} and to eigenvalue optimization problems \cite{Bu06}.
We did not even consider the effect of nonlinearites. We mention the closely related effect
of discontinuous change of the critical flutter frequency due to small dissipation \cite{BZ69,KS05b} and its connection to
the Whitney umbrella singularity at the exceptional points on the eigenvalue surfaces \cite{SKM05}.
Another related topic is the role of the spectral exceptional points in modern non-Hermitian physics
including crystal optics, open quantum systems, and $\mathcal{PT}$-symmetric quantum mechanics \cite{B04}.
All this shows that modern non-conservative and non-Hermitian problems are a perfect field of applied
mathematics with a big potential for new discoveries.

\section*{Acknowledgements}
The work of O.N.K. has been supported by the research grant DFG HA 1060/43-1.


\begin{thebibliography}{99}

\bibitem{AY75}
I.P. Andreichikov, V.I. Yudovich, {\em The stability of
visco-elastic rods}, Izv. Acad. Nauk SSSR. MTT, 1 (1975),
150--154.

\bibitem{AA88}
D.V. Anosov and V.I. Arnold (eds.), Dynamical Systems I, Encyclopaedia of Mathematical Sciences,
Springer, Berlin etc. 1988.

\bibitem{Ar71}
V.I. Arnold, {\em On matrices depending on parameters},
Russian Math. Surveys., 26 (1971), 29--43.

\bibitem{Ar83}
V.I. Arnold, {\em Geometrical Methods in the Theory of Ordinary Differential Equations},
New York: Springer-Verlag, 1983.

\bibitem{AA93}
V.I. Arnold (ed.), Dynamical Systems VIII, Encyclopaedia of Mathematical Sciences, Springer, Berlin etc.
1993.

\bibitem{BBM89a}
N.V. Banichuk, A.S. Bratus, A.D. Myshkis,
{\em On destabilizing influence of small dissipative forces to nonconservative
systems}, Doklady AN SSSR 309(6) (1989), 1325--1327.

\bibitem{BBM89}
N.V. Banichuk, A.S. Bratus, A.D. Myshkis, {\em Stabilizing
and destabilizing effects in nonconservative systems}, {PMM
U.S.S.R.}, {53}(2) (1989), 158--164.

\bibitem{B04}
M.V. Berry,
{\em Physics of non-Hermitian degeneracies},
Czech. J. Phys. {54} (2004), 1039--1047.

\bibitem{Br94}
A.M. Bloch, P.S. Krishnaprasad, J.E. Marsden, T.S. Ratiu,
{\em Dissipation-induced instabilities}, Annales de l'Institut Henri
$\rm Poincar\acute{e}$, {11}(1) (1994), 37--90.

\bibitem{Bo63}
V.V. Bolotin, {\em Non-conservative Problems of the Theory of
Elastic Stability}, Fizmatgiz (in Russian), Moscow, 1961; Pergamon, Oxford, 1963.

\bibitem{BZ69}
V.V. Bolotin, N.I. Zhinzher, {\em Effects of damping on stability of elastic systems subjected to
nonconservative forces}, Int. J. Solids Struct. 5 (1969), 965--989.

\bibitem{Bo02}
V.V. Bolotin, A.A. Grishko, M.Yu. Panov, {\em Effect of damping on the
postcritical behavior of autonomous non-conservative systems},
{Intern. J. of Nonl. Mechs.} {37} (2002), 1163--1179.

\bibitem{B55}
O. Bottema, {\em On the stability of the equilibrium of a linear mechanical system},
Z. Angew. Math. Phys. 6 (1955), 97--104.

\bibitem{B56}
O. Bottema,
{\em The Routh-Hurwitz condition for the biquadratic equation},
Indagationes Mathematicae, 18 (1956), 403--406.

\bibitem{B08}
N. M. Bou-Rabee, J. E. Marsden, L. A. Romero, {\em Dissipation-Induced
Heteroclinic Orbits in Tippe Tops},
SIAM Review, 50(2) (2008), 325--344.

\bibitem{Bu06}
J.V. Burke, D. Henrion, A.S. Lewis, M.L. Overton, {\em Stabilization via Nonsmooth, Nonconvex
Optimization}, IEEE Transactions on Automatic Control, 51(11) (2006), 1760--1769.


\bibitem{CM79}
J. Carr, M.Z.M. Malhardeen, {\em Beck's problem},
SIAM J. Appl. Math. 37(2) (1979), 261--262.

\bibitem{CE49}
J.G. Charney, A. Eliassen, {\em A numerical method for predicting the perturbations of
the middle latitude westerlies}, Tellus 1 (1949) 38--54.

\bibitem{C95}
S.H. Crandall, {\em The effect of damping on the stability of gyroscopic
pendulums.} Z. angew. Math. Phys. 46 (1995), S761--S780.

\bibitem{E49}
E.T. Eady, {\em Long waves and cyclone waves}, Tellus 1 (1949) 38--52.

\bibitem{GS85}
M. Golubitsky and D.G. Schaeffer, Singularities and maps in bifurcation theory, vol. 1, Applied Mathematical
Sciences 51, Springer, Berlin etc. 1985.

\bibitem{GS88}
M. Golubitsky, D.G. Schaeffer and I. Stewart, Singularities and maps in bifurcation theory, vol. 2, Applied
Mathematical Sciences 69, Springer, Berlin etc. 1988.

\bibitem{Ha92}
G. Haller, {\em Gyroscopic stability and its loss in systems
with two essential coordinates}, Intern. J. of Nonl. Mechs., 27
(1992), 113--127.

\bibitem{Ha75}
W. Hauger,
{\em Stability of a gyroscopic non-conservative system},
{Trans. ASME, J. Appl. Mech.}, 42 (1975), 739--740.

\bibitem{HJ65}
G. Herrmann and I.~C. Jong, {\em On the destabilizing effect
of damping in nonconservative elastic systems}, ASME J. of Appl.
Mechs., {32}(3) (1965), 592--597.

\bibitem{H67}
G. Herrmann, {\em Stability of equilibrium of elastic systems subjected to non-conservative forces},
Appl. Mech. Revs. 20 (1967), 103--108.

\bibitem{HD92}
K. Higuchi, E.H. Dowell,
{\em Effect of structural damping on flutter of plates with a follower force},
AIAA J. 30(3) (1992), 820--825.

\bibitem{HG03}
N. Hoffmann, L. Gaul,
{\em Effects of damping on mode-coupling instability in friction
induced oscillations},
Z. angew. Math. Mech., 83 (2003), 524--534.

\bibitem{Ho61}
E.O. Holopainen,
{\em On the effect of friction in baroclinic waves},
Tellus. 13(3) (1961), 363--367.

\bibitem{HR95}
I. Hoveijn, M. Ruijgrok, {\em The stability of
parametrically forced coupled oscillators in sum resonance},
Z. angew. Math. Phys.,
46 (1995), 384--392.

\bibitem{H93}
J. ${\rm Hult\acute{e}n}$, {\em Drum brake squeal---a self-exciting mechanism with constant friction.}
In: SAE Truck and Bus Meeting, 1993, Detroit, Mi, USA, SAE Paper 932965.

\bibitem{IA92}
G. Iooss, M. Adelmeyer, {\em Topics in bifurcation theory},
World Scientific, Singapore (1992).

\bibitem{KOS07}
D.V. Kapitanov, V.F. Ovchinnikov, L.V. Smirnov,
{\em Numerical-analytical stability investigation of beam with servo force fixed as cantilever at free end}, Problems of Strength and Plasticity. 69 (2007), 177--184.

\bibitem{K39}
P.L. Kapitsa, {\em Stability and passage through the critical speed of the fast spinning rotors in the presence of damping}, Zh. Tech. Phys. 9(2) (1939), 124--147.

\bibitem{KZ07}
P. Kessler, O.M. O'Reilly, A.-L. Raphael, M. Zworski, {\em On dissipation-induced
destabilization and brake squeal: a perspective using structured pseudospectra}, J. Sound Vib.
308 (2007), 1--11.

\bibitem{K24}
A.L. Kimball, {\em Internal friction theory of shaft whirling}, Gen. Elec. Rev. 27 (1924), 224--251.

\bibitem{K03a}
O.N. Kirillov, {\em How do small velocity-dependent forces (de)stabilize a non-conservative system?}
DCAMM Report. No. 681. April 2003. 40 pages.

\bibitem{K03b}
O.N. Kirillov, {\em How do small velocity-dependent forces (de)stabilize a non-conservative system?}
Proceedings of the International Conference "Physics and Control". St.-Petersburg. Russia.
August 20-22. 2003. Vol. 4, 1090--1095.

\bibitem{Ki04}
O.N. Kirillov, {\em Destabilization paradox}, Doklady Physics, 49(4) (2004), 239--245.

\bibitem{K05}
O.N. Kirillov, {\em A theory of the destabilization paradox in non-conservative systems},
Acta Mechanica. 174(3-4) (2005), 145--166.

\bibitem{KS05a}
O.N. Kirillov, A.P. Seyranian, {\em Stabilization and destabilization of a circulatory system by small velocity-dependent forces}, J. Sound Vibr., 283(3-5) (2005), 781--800.

\bibitem{KS05b}
O.N. Kirillov, A.P. Seyranian, {\em The effect of small internal and external damping on the stability of distributed non-conservative systems}, J. Appl. Math. Mech. 69(4) (2005), 529--552.

\bibitem{KS05c}
O.N. Kirillov, A.P. Seyranian, {\em Instability of distributed nonconservative systems caused by weak dissipation},
Doklady Mathematics. 71(3) (2005), 470--475.

\bibitem{K06}
O.N. Kirillov, {\em Gyroscopic stabilization of non-conservative systems},
Phys. Lett. A. 359(3) (2006), 204--210.

\bibitem{K07a}
O.N. Kirillov, {\em Destabilization paradox due to breaking the Hamiltonian and reversible symmetry},
Int. J. Non-Lin. Mech. 42(1) (2007), 71--87.

\bibitem{K07b}
O.N. Kirillov, {\em Gyroscopic stabilization in the presence of nonconservative forces},
Dokl. Math. 76(2) (2007), 780--785.

\bibitem{K07c}
O.N. Kirillov, {\em Bifurcation of the roots of the characteristic polynomial and destabilization paradox in friction induced oscillations}, Theor. Appl. Mech. 34(2) (2007), 87--109.

\bibitem{K08}
O.N. Kirillov, {\em Subcritical flutter in the acoustics of friction},
Proc. R. Soc. A  464(2097) (2008), 2321--2339.

\bibitem{K08p}
O.N. Kirillov, Perturbation of multiparameter non-self-adjoint boundary eigenvalue problems for operator matrices.
Preprint arXiv:0803.2248v2 [math-ph] 14 Mar 2008.

\bibitem{K09}
O.N. Kirillov, Sensitivity analysis of Hamiltonian and reversible systems prone to dissipation-induced instabilities. in: {\em Matrix methods: theory, algorithms, applications}, E. Tyrtyshnikov and V. Olshevsky, eds. World Scientific. 2009. P. 31--68.

\bibitem{Ko92}
A.N. Kounadis, {\em On the paradox of the destabilizing
effect of damping in nonconservative systems}, Intern. J. of Nonl.
Mechs., {27} (1992), 597--609.


\bibitem{KM06}
R. Krechetnikov, J.E. Marsden,
{\em On destabilizing effects of two fundamental non-conservative forces},
Physica D, 214 (2006), 25--32.

\bibitem{KM07}
R. Krechetnikov, J.E. Marsden,
{\em Dissipation-induced instabilities in finite dimensions},
Rev. Mod. Phys. 79 (2007), 519--553.

\bibitem{KM09}
R. Krechetnikov, J.E. Marsden, {\em Dissipation-Induced Instability Phenomena in Infinite-Dimensional Systems},
Arch. Rat. Mech. Anal. (2009). DOI 10.1007/s00205-008-0193-6



\bibitem{Kuz04}
Yu.A. Kuznetsov, Elements of applied bifurcation theory, Applied Mathematical Sciences 112, Springer,
Berlin etc. 2004.

\bibitem{La03}
W.F. Langford, {\em Hopf meets Hamilton under Whitney's umbrella}, in
IUTAM symposium on nonlinear stochastic dynamics.
Proceedings of the IUTAM symposium, Monticello, IL, USA, Augsut 26-30,
2002, Solid Mech. Appl. 110, S.~N. Namachchivaya, et al., eds., Kluwer, Dordrecht, 2003, pp.~157--165.

\bibitem{L80}
L.V. Levantovskii,  {\em The boundary of a set of stable matrices},
Uspekhi Mat. Nauk 35 (1980), no. 2(212), 213--214.

\bibitem{L82}
L.V. Levantovskii, {\em Singularities of the boundary of a region of stability}, (Russian)
Funktsional. Anal. i Prilozhen. 16 (1982), no. 1, 44--48, 96.

\bibitem{MK86}
R.S. MacKay,
Stability of equilibria of Hamiltonian systems. In \textit{Nonlinear Phenomena and Chaos} (ed. S. Sarkar),
Adam Hilger, Bristol, (1986), 254--270.

\bibitem{MK91}
R.S. MacKay, {\em Movement of eigenvalues of Hamiltonian
equilibria under non-Hamiltonian perturbation}, Phys. Lett. A,
{155} (1991), 266--268.

\bibitem{MO95}
J. Maddocks, M.L. Overton, {\em Stability theory for
dissipatively perturbed Hamiltonian systems}, Comm. Pure and
Applied Math., {48} (1995), 583--610.

\bibitem{OR96}
O.M. O'Reilly, N.K. Malhotra, N.S. Namachchivaya, {\em Some aspects of
destabilization in reversible dynamical systems with application to
follower forces}, Nonlin. Dyn. {10} (1996), 63--87.

\bibitem{PS87}
Ya.G. Panovko, S.V. Sorokin, {On quasi-stability of viscoelastic systems with the follower forces},
Izv. Acad. Nauk SSSR. Mekh. Tverd. Tela. 5 (1987), 135--139.

\bibitem{PP02}
K. Popp, M. Rudolph, M. Kr\"oger, M. Lindner,
{\em Mechanisms to generate and to avoid friction induced vibrations},
VDI-Berichte {1736}, VDI-Verlag, $\rm D\ddot usseldorf$, 2002.

\bibitem{R77}
R.A. Romea, {\em The effects of friction and $\beta$ on finite-amplitude baroclinic waves},
J. Atmos. Sci. 34 (1977), 1689--1695.

\bibitem{RTV93}
M. Ruijgrok, A. Tondl, F. Verhulst, {\em Resonance in a Rigid Rotor with Elastic Support},
Z. angew. Math. Mech. 73 (1993), 255--263.

\bibitem{Sa08}
A.K. Samantaray, R. Bhattacharyya, A. Mukherjee,
{\em On the stability of Crandall gyropendulum}, Phys. Lett. A 372 (2008), 238--243

\bibitem{S04}
V.A. Samsonov, T.S. Sumin,
{\em On the stability of the equilibrium position of a mechanical system with two degrees of freedom},
Vestnik Moskov. Univ. Ser. I Mat. Mekh., 4 (2004), 60--62, 72.

\bibitem{SVM}
J.A. Sanders,  F. Verhulst, J. Murdock,
{\em Averaging methods in nonlinear dynamical systems},
Applied Math. Sciences 59, Springer (2007, rev. ed.).

\bibitem{Se90}
A.P. Seyranian, {\em Destabilization paradox in stability
problems of non-conservative systems}, Advances in Mechanics,
{13}(2) (1990), 89--124.

\bibitem{S96}
A.P. Seyranian, {\em On stabilization of non-conservative systems by dissipative forces and
uncertainty of critical load}, Doklady Akademii Nauk. 348 (1996), 323--326.

\bibitem{SKM05}
A.P. Seyranian, O.N. Kirillov, A.A. Mailybaev,
{\em Coupling of eigenvalues of complex matrices at diabolic and exceptional points},
J. Phys. A: Math. Gen. 38(8) (2005), 1723--1740.

\bibitem{Sey}
A.P. Seyranian,  A.A. Mailybaev,
{\em Multiparameter stability theory with mechanical applications},
World Scientific, series A, vol. 13 (2003).

\bibitem{SK03}
A.P. Seyranian, O.N. Kirillov, {\em Effect of small dissipative and gyroscopic forces on the stability of nonconservative systems}, Doklady Physics, 48(12) (2003), 679--684.

\bibitem{SJ07}
J.-J. Sinou, L. Jezequel,
{\em Mode coupling instability in friction-induced vibrations and its dependency on system parameters including damping},
Eur. J. Mech. A. 26 (2007), 106--122.

\bibitem{S33}
D.M. Smith, {\em The motion of a rotor carried by a flexible shaft in flexible
bearings}, Proc. Roy. Soc. London A 142 (1933), 92--118.

\bibitem{S08}
G. Spelsberg-Korspeter, O.N. Kirillov, P. Hagedorn,
{\em Modeling and stability analysis of an axially moving beam with frictional contact},
Trans. ASME, J. Appl. Mech. 75(3) (2008), 031001.

\bibitem{SS90}
W. Szemplinska-Stupnicka, {\em The behaviour of nonlinear vibrating systems}, Vol. II,
Kluwer, Dordrecht etc. (1990).

\bibitem{TT79}
W. Thomson, P.G. Tait, {\em Treatise on Natural Philosophy}, Vol. I, Part I, New Edition, pp.
387-391, Cambridge Univ. Press, Cambridge (1879).

\bibitem{T95}
J.J. Thomsen, {\em Chaotic dynamics of the partially follower-loaded elastic double pendulum},
J. Sound Vibr. 188(3) (1995), 385--405.

\bibitem{TZ81}
H. Troger, K. Zeman, {\em Zur korrekten Modellbildung in der Dynamik diskreter Systeme}, Ing.-Arch. 51 (1981), 31--43.

\bibitem{V02}
F. Verhulst,
{\em Parametric and Autoparametric Resonance},
Acta Appl. Math., 70(1-3) (2002), 231--264.

\bibitem{V09}
F. Verhulst. Perturbation analysis of parametric resonance, Encyclopedia of Complexity and Systems Science, Springer, 2009.

\bibitem{W43}
H. Whitney, {\em The general type of singularity of a set of $2n - 1$ smooth functions of n
variables}, Duke Math. J., 10 (1943), 161--172.

\bibitem{W44}
H. Whitney,
{\em The singularities of a smooth $n$-manifold in $(2n - 1)$-space}, Ann. of Math.,
45(2) (1944), 247--293.


\bibitem{YS75}
V.A. Yakubovich, V.M. Starzhinskii, {\em Linear differential equations with periodic coefficients}, 2 vols.,
John Wiley, New York etc. (1975).

\bibitem{Z07}
Z.Y. Zhang, J. Neubauer, D.A. Berry,
{\em Physical mechanisms of phonation onset: A linear stability analysis of an aeroelastic continuum model of phonation.}
J. Acoust. Soc. of Amer., 122(4) (2007), 2279--2295.

\bibitem{Z94}
N.I. Zhinzher, {\em Effect of dissipative forces with incomplete dissipation on the stability of elastic systems}, Izv. Ross. Acad. Nauk. MTT 19 (1994), 149--155.

\bibitem{Zhu92}
V.F. Zhuravlev, {\em Nutational vibrations of a free
gyroscope}, Izv. Ross. Akad. Nauk, Mekh. Tverd. Tela, {6} (1992),
13--16.

\bibitem{Zi52}
H. Ziegler, {\em Die Stabilit\"atskriterien der Elastomechanik},
Ing.-Arch., 20 (1952), 49--56.

\bibitem{Zi53}
H. Ziegler, {\em Linear elastic stability: A critical analysis of methods},
Z. Angew. Math. Phys, 4 (1953), 89--121.

\end{thebibliography}
\end{document}